\documentclass[%
 reprint,
 amsmath,amssymb,
 aps,
 onecolumn
]{revtex4-2}

\usepackage{graphicx}% Include figure files
\usepackage[outdir=./]{epstopdf}
\usepackage{tikz}
\usetikzlibrary{arrows.meta,decorations.markings}
\usepackage[toc,page]{appendix}
\usepackage{xcolor,amsmath,multirow,mathrsfs}
\usepackage{dcolumn}% Align table columns on decimal point
\usepackage {lipsum}
\usepackage{tikz}
\usetikzlibrary{arrows.meta}
\usepackage{bm}% bold math
\usepackage{hyperref}% add hypertext capabilities
\hypersetup{
	colorlinks=true,
	linkcolor=magenta,
	filecolor=magenta,    
	citecolor=red,  
	urlcolor=cyan,
}
\allowdisplaybreaks
\usepackage[mathlines]{lineno}% Enable numbering of text and display math
\usepackage{float}
\usepackage{fullpage,tikz}
\usetikzlibrary{arrows.meta,decorations.markings,patterns}
\restylefloat{table}
\usepackage{multirow}
\newcommand\numberthis{\addtocounter{equation}{1}\tag{\theequation}}
\usepackage{ifthen}
\newboolean{showcomments}
\setboolean{showcomments}{true}

\newcommand{\dennice}[1]{\ifthenelse{\boolean{showcomments}}
{\textcolor{blue}{Dennice says: #1}}{}}
\usepackage{enumitem}
\usepackage[utf8]{inputenc}
	\definecolor{applegreen}{rgb}{0.55, 0.71, 0.0}
	\definecolor{ao}{rgb}{0.0, 0.5, 0.0}
\newcommand{\blueline}{\raisebox{2pt}{\tikz{\draw[-,blue,solid,line width = 1.0pt](0,0) -- (5mm,0);}}}
\newcommand{\bluedashedline}{\raisebox{2pt}{\tikz{\draw[-,blue,dash pattern={on 6pt off 3pt on 6pt},line width = 1.0pt](0,0) -- (5mm,0);}}}

\newcommand{\redline}{\raisebox{2pt}{\tikz{\draw[-,red,solid,line width = 1.0pt](0,0) -- (5mm,0);}}}

\newcommand{\greendashdotline}{\raisebox{2pt}{\tikz{\draw[-,ao,dash dot,line width = 1.0pt](0,0) -- (5mm,0);}}}

\newcommand{\blackline}{\raisebox{2pt}{\tikz{\draw[-,black,solid,line width = 1.0pt](0,0) -- (5mm,0);}}}

\newcommand{\tikzcircle}[2][red,fill=red]{\tikz[baseline=-0.5ex]\draw[#1,radius=#2] (0,0) circle ;}%

\begin{document}
\preprint{APS/123-QED}

\title{Effects of wind veer on a yawed wind turbine wake \\ in atmospheric boundary layer flow}% Force line breaks with \\
%\thanks{A footnote to the article title}%

\author{Ghanesh\,Narasimhan, Dennice\,F.\,Gayme and Charles\,Meneveau}
 %\altaffiliation[Also at ]{Physics Department, XYZ University.}%Lines break automatically or can be forced with \\
%\author{Second Author}%
 %\email{Second.Author@institution.edu}
\affiliation{Department of Mechanical Engineering, Johns Hopkins University, Baltimore, Maryland 21218, USA}

%\collaboration{MUSO Collaboration}%\noaffiliation

%\author{Charlie Author}
% \homepage{http://www.Second.institution.edu/~Charlie.Author}
%\affiliation{
 %Second institution and/or address\\
 %This line break forced% with \\
%}%
%\affiliation{
 %Third institution, the second for Charlie Author
%}%
%\author{Delta Author}
%\affiliation{%
 %Authors' institution and/or address\\
 %This line break forced with \textbackslash\textbackslash
%}%

%\collaboration{CLEO Collaboration}%\noaffiliation

%\date{\today}% It is always \today, today,
             %  but any date may be explicitly specified

\begin{abstract}
% Yawing wind turbines can increase wind farm power output potential by steering the wake away from downstream turbines.  The spanwise aerodynamic force due to yaw is known to generate a counter-rotating vortex pair (CVP) at the top and bottom of the turbine. This CVP induces a side-wash velocity that causes deflection and curling of the wake, which affects the power output of downstream turbines.  This work uses Large Eddy Simulation (LES) to investigate how wind veer in a Conventionally Neutral Atmospheric Boundary Layer (CNBL) affects the wake evolution and properties of the CVP arising from turbine yaw. LES of a single yawed wind turbine in a CNBL with geostrophic forcing indicates that the presence of wind veer contributes to an asymmetry in the circulation strengths of the CVP's top and bottom vortices. These simulations are used to inform a simple correction [see Abkar et al, Energies (2018), 11(7), 1838] for traditional models that do not explicitly include veer. Comparisons between LES and corrections to existing analytical models for yawing turbines indicate good agreement in the downstream decay of maximum vortex-core vorticity, the vortices' circulation strength, and the curled wake shape. 

Large Eddy Simulations (LES) are used to study the effects of veer (the height-dependent lateral deflection  of wind velocity due to Coriolis acceleration) on the evolution of wind turbine wakes. Specifically, this work focuses on turbines that are yawed with respect to the mean incoming wind velocity, which produces laterally deflected wakes that have a curled (crescent-shaped) structure. These effects can be attributed to the introduction of streamwise mean vorticity and the formation of a Counter-rotating Vortex Pair (CVP) on the top and bottom of the wake. In a {\color{black} Truly Neutral Boundary Layer (TNBL)} in which wind veer effects are absent, these effects can be captured well with existing analytical wake models (Bastankhah et al. J. Fluid Mech. (2022), 933, A2). However, in the more realistic case of atmospheric boundary layers subjected to Coriolis acceleration, existing models need to be re-examined and generalized to include the effects of wind veer. To this end, the flow in a Conventionally Neutral Atmospheric Boundary Layer (CNBL) interacting with a yawed wind turbine is {\color{black} investigated in this study}. Results indicate that in the presence of veer the CVP's top and bottom vortices exhibit considerable asymmetry. However, upon removing the veer component of vorticity, the resulting distribution is much more symmetric and agrees well with that observed in a TNBL. These results are used to develop a simple correction to predict the mean velocity distribution in the wake of a yawing turbine in a CNBL using analytical models. The correction includes the veer-induced sideways wake deformation, as proposed by Abkar et al. (Energies (2018), 11(7), 1838). The resulting model predictions are compared to mean velocity distributions from the LES and good agreement is obtained.

\end{abstract}
%subtracting the wind veer component of vorticity due to mean flow
%\keywords{Suggested keywords}%Use showkeys class option if keyword
                              %display desired
\maketitle
%\tableofcontents
\section{Introduction}\label{Intro}
Yawing a turbine deflects its wake, decreasing wake interactions and potentially increasing the power output of downstream turbines \cite{fleming2019}. Coordinating such actions over a wind farm could improve its overall efficiency  \cite{howland19}. 
% The yaw-induced wake deflection has been attributed to the formation of a Counter-rotating Vortex Pair (CVP) \citep{bastankhah2016,howland16,shapiro2018}. This CVP is generated through the force exerted by yawing the turbine on the passing air, which has a spanwise component that varies in the spanwise direction, thus creating vorticity and generating the CVP on the top and bottom of the turbine rotor \cite{shapiro2020}. These vortices induce a side wash velocity on the wake that deforms and deflects it away from the center of the turbine \cite{howland16}.
%\dennice{Needs a connecting thought that says there remains more to be known and then this sentence. } 
Wake deflection due to yaw was studied experimentally  \citep{Grant_et_al_1997,Grant_Parkin_2000,Parkin_et_al_2001,Wouter_et_al_2005,Medici_Alfredsson_2006}, while Ref. \cite{jimenez2010} performed an early Large Eddy Simulation (LES) study and proposed a simple analytical model for predicting the initial wake skewing angle just behind the turbine. In a subsequent wind tunnel study \cite{howland16}, the formation of an axial Counter-rotating Vortex Pair (CVP) was observed behind a yawed actuator disk, in the presence of a uniform inflow. The deflection of the wake was attributed to the CVP because the vortices (one above and the other below the actuator disk) induce a side wash velocity that deflects the wake from the center of the turbine. The vortices also deform the wake shape into a curled (crescent-shaped) structure.  Ref. \cite{bastankhah2016} performed further wind tunnel studies of a model wind turbine in a turbulent boundary layer where the CVP formation was also observed. 
%Ref. \cite{bastankhah2016} developed an analytical model to quantify the downstream wake deflection based on Reynolds-averaged Navier Stokes (RANS) equation. 

Ref. \cite{shapiro2018}  proposed considering the turbine as a lifting surface (applying a height-dependent sideways force onto the fluid), i.e., analogous to a vertically placed airfoil that sheds streamwise (tip) vortices in the presence of an incoming mean flow. Evaluation of the induced strength of the CVP near the turbine enabled predicting the yaw-induced wake deflection quite accurately \cite{shapiro2018}.  Other vortex-based models describe the vorticity at the turbine as a distribution of multiple, discrete point-vortices \citep{Martinez_et_al_2020,Martinez_Tossas_2020,zong_porte_2020}   whose downstream transport  and diffusion are modeled numerically.  Following these studies, Ref. \cite{shapiro2020} proposed a theory for the generation and downstream evolution of the CVP. The analytical predictions for the decay of the maximum vorticity and circulation strength of the vortices showed very good agreement with the LES data, while still assuming a circular shape of the wake. 
% In a more detailed recent study Ref. \cite{bastankhah_et_al_2022}, using the CVP's circulation  strength estimate from Ref. \cite{shapiro2020}, it is further shown that an analytical model can be derived that successfully predicts the curled wake shape behind yawed turbines.  
In a more detailed recent study \cite{bastankhah_et_al_2022}, it was shown that an analytical vortex sheet-based model can successfully predict the curled wake shape behind yawed turbines. In this model, the wake edge was treated as a vortex sheet and analytical solutions using truncated power series expansions were obtained based on the decaying circulation strength estimate of the CVP from Ref. \cite{shapiro2020}. The Gaussian wake model for the axial velocity deficit in \cite{bastankhah2016} was then modified to include the deformation caused by the vortex sheet that predicted the curled shape and the deflection of the  wake quite accurately.   

Wind turbine wake properties and the performance of wind farms also depend on the prevailing properties of the atmospheric boundary layer (ABL). For instance, it is well known that the wake recovery rate (i.e., the wake expansion coefficient) is affected by the ABL's thermal stratification conditions \cite{Abkar_Porteagel_2015}. At the same time, the Coriolis acceleration due to Earth's rotation causes an Ekman spiral flow in the surface layer of the ABL \cite{ekman1905}. This leads to a height-dependent lateral realignment of the incoming wind direction called wind veer, which can significantly affect the wind farm power output \cite{Gadde_2019}.  Veer effects have been previously considered for an unyawed turbine in a stably stratified ABL \cite{Abkar_et_al_2018}. That work modified the Gaussian wake model with a veer correction term which successfully predicted the skewed/sheared wake structure arising from the spanwise shear due to the wind veer. Ref. \cite{howland_et_al_2020} considered the effects of both yaw and veer on individual turbine blade aerodynamics.  However, the combined effects of veer and yawing on wind turbine wakes and their modeling via analytical approaches have not received significant attention so far. The objective of the present study is to examine the evolution of turbine wakes in the presence of both turbines yawing and veer, as well as to include both of these effects in analytical models of the turbine wake.

In order to generate the relevant data, an LES of a yawed wind turbine in the presence of an incoming mean flow typical of a Conventionally-Neutral ABL (CNBL) that includes veer is performed. The CNBL is a type of ABL characterized by a neutrally stratified turbulent boundary layer region separated from the Geostrophic and stably stratified free atmosphere by a capping inversion layer at the boundary layer height \cite{Allaerts_Meyers_2015}.  As a reference, we also perform the LES of a yawed wind turbine in a {\color{black} truly neutral boundary layer (TNBL)}.  The details of the LES are described in \S \ref{LES}.  The results are analyzed in section \ref{veer_vort_dyn}, which focuses on the effect of wind veer on the downstream evolution of mean vorticity.  The results are used to introduce modifications to existing analytical models so that both ABL veer and turbine yaw can be represented efficiently and accurately as described in \S \ref{vortex_sheet_model}. The main conclusions are summarized in \S \ref{conclusion}.

\section{Large Eddy Simulation of a yawed wind turbine in a CNBL }\label{LES} 

This section details the LES setup of {\color{black} CNBL} and {\color{black} TNBL} simulations used to generate data to study the effect of wind veer on a yawed turbine wake. 
We use the open-source code LESGO \cite{LESGO}, an LES solver  primarily developed to simulate ABL flows \cite{Albertson_1999,bouzeid2005}. The code includes various dynamic sub-grid stress parameterizations, wall models, wind turbine representations using actuator disk/line models, and inflow generation using the concurrent-precursor approach \cite{stevens2014}. The code has been validated by several previous studies \citep{calaf_et_al_2010,Calaf_et_al_2011,stevens2014,stevens_et_al_JRSE_2014,martinez_et_al_2015,bouzeid2005,shapiro_et_al_AD_2019,shapiro2018,shapiro2020}.
The governing equations and numerical method, initial conditions for the velocity, and potential temperature are discussed in \S \ref{gov_eqn_nm_sec}. 
The simulation setup for the current LES study is described in \S \ref{sim_setup_sec}.
The characteristics of the CNBL and TNBL to be simulated are described in \S \ref{prec_result_sec}, which documents the main differences in mean flow velocity profiles between both cases.

\subsection{Governing equations and numerical method \label{gov_eqn_nm_sec}}
The code LESGO solves the filtered Navier-Stokes equations (with the Boussinesq approximation for buoyancy effects) and the scalar potential temperature transport equation: 
%\dennice{Shouldn't we include the equations here particularly since we are including additional terms for the CNBL}. 
%The corresponding equations are
\begin{align*}
    \frac{\partial \tilde{u}_i}{\partial x_i}&=0,\numberthis\label{continuity_LES}\\
    \frac{\partial \tilde{u}_i}{\partial t}+\tilde{u}_j\left(\frac{\partial \tilde{u}_i}{\partial x_j}-\frac{\partial \tilde{u}_j}{\partial x_i}\right)&=-\frac{1}{\rho_0}\frac{\partial p_\infty}{\partial x_i} -\frac{\partial \tilde{p}}{\partial x_i}+\frac{g}{\tilde{\theta}_0}(\tilde{\theta}-\tilde{\theta}_0)\delta_{i3}-\frac{\partial\tau_{ij}}{\partial x_j}\\
    &\mspace{20mu}+\frac{1}{\rho_0}\tilde{f}_{x}\ \delta_{i1}+\frac{1}{\rho_0}\tilde{f}_{y}\ \delta_{i2}-f_c\tilde{u} \ \delta_{i2}+f_c\tilde{v} \ \delta_{i1},\numberthis\label{momentum_LES}\\
    \frac{\partial\tilde{\theta}}{\partial t}+\tilde{u}_j\frac{\partial \tilde{\theta}}{\partial x_j}&=-\frac{\partial \Pi_j}{\partial x_j},\numberthis\label{theta_LES}
\end{align*}
where the tilde ($\tilde{\cdot}$) represents spatial filtering operation such that $\tilde{u}_i=(\tilde{u},\tilde{v},\tilde{w})$ are the filtered velocity components in the streamwise, lateral and vertical directions, respectively, and $\tilde{\theta}$ is the filtered potential temperature. The term $\tau_{ij}=\sigma_{ij}-(1/3)\sigma_{kk}\delta_{ij}$ is the deviatoric part of the Sub-Grid Scale (SGS) stress tensor $\sigma_{ij}=\widetilde{u_i u_j}-\tilde{u}_i\tilde{u}_j$. The quantity $\tilde{p}=\tilde{p}_*/\rho_0+(1/3)\sigma_{kk}+(1/2)\tilde{u}_j\tilde{u}_j$ is the modified pressure, where the actual pressure $\tilde{p}_*$ divided by the ambient density $\rho_0$ is augmented with the trace of the SGS stress tensor and the kinematic pressure arising from writing the non-linear terms in rotational form. The quantities $\tilde{f}_i=(\tilde{f}_x,\tilde{f}_y,0)$ are the streamwise and spanwise component of the turbine's force imparted on the fluid. The term $-(1/\rho_0)\partial p_\infty/\partial x_i$ is the mean external pressure gradient applied to drive the flow. The $\delta_{ij}$ in equation \eqref{momentum_LES} is the Kronecker delta function determining the direction of buoyancy, turbine thrust and Coriolis forces. In the buoyancy term, $g=9.8 \ \text{m}/\text{s}^2$ is the gravitational acceleration, $\tilde{\theta}_0$ is the reference potential temperature scale taken to be 288 K for the CNBL case. In the Coriolis force terms, $f_c=2\Omega\sin\phi=10^{-4} \ \text{s}^{-1}$ is the Coriolis parameter at latitude $\phi=45^\circ$. In equation \eqref{theta_LES}, the term $\Pi_j=\widetilde{u_j \theta}-\tilde{u}_j\tilde{\theta}$ is the SGS heat flux.

In the momentum equation \eqref{momentum_LES}, a constant mean pressure gradient $\nabla p_\infty=(\partial p_\infty/\partial x,\partial p_\infty/\partial y,0)$ is applied to drive the flow. For the CNBL case, it is written in terms of the Geostrophic velocity components $U_g,V_g$ using the Geostrophic balance equation
\begin{align}
    \frac{1}{\rho_0}\frac{\partial p_\infty}{\partial x}= f_c V_g,\quad \frac{1}{\rho_0}\frac{\partial p_\infty}{\partial y}=-f_c U_g.\label{dpdx_ext}
\end{align}
The Geostrophic velocity components are specified as $U_g=G\cos\alpha, V_g=G\sin\alpha$, where  $G=(U_g^2+V_g^2)^{1/2}$ is the magnitude of the Geostrophic wind which is set to 8 m/s and $\alpha$, is the angle made by the resultant wind vector with respect to the streamwise $x$ direction. For the TNBL case, the dimensionless streamwise mean pressure gradient is set to a constant value. {\color{black} This constant streamwise pressure gradient ensures the mean flow is streamwise aligned throughout the domain without any wind veer ($V(z)\equiv 0$).} In addition, since the TNBL flow is isothermal and neutrally buoyant throughout the domain, equation \eqref{theta_LES} is not solved for this case, and also the buoyancy term in the momentum equation \eqref{momentum_LES} vanishes. 

In the CNBL simulation, we maintain a mean flow direction such that it is streamwise aligned at the hub height. This is achieved by choosing a value of $\alpha$ for the Geostrophic wind such that the wind veer at hub height is zero $(V(z=0)=0)$. To compute the appropriate value of $\alpha$, we use the  Proportional-Integral (PI) control approach introduced in Ref. \cite{sescu2014} with a proportional gain $K_P=10$, and an integral gain $K_I=0.5$. We also impose the constraint $V(z=0)=0$.

We solve the equations for the high Reynolds number limit such that the molecular viscous and heat diffusion terms are neglected in the equations \eqref{momentum_LES} and \eqref{theta_LES}. The necessary diffusion for the problem is provided by modeling the deviatoric part of the SGS stress tensor $(\tau_{ij})$ and the SGS heat flux $(\Pi_j)$ as
\begin{align}
    \tau_{ij}&=-2\nu^{\text{SGS}}_T\tilde{S}_{ij}, \mspace{50mu}
    \Pi_j=-\kappa^{\text{SGS}}_T\frac{\partial\tilde{\theta}}{\partial x_j},\label{SGS_model}
\end{align}
where $\nu^{\text{SGS}}_T$ is the SGS momentum diffusivity, $\kappa^{\text{SGS}}_T$ is the SGS heat diffusivity, $\tilde{S}_{ij}=(1/2)(\partial \tilde{u}_i/\partial x_j+\partial \tilde{u}_j/\partial x_i)$ is the symmetric part of the velocity gradient tensor. The diffusivities are related by the SGS Prandtl number $Pr_{\text{SGS}}=\nu^{\text{SGS}}_T/\kappa^{\text{SGS}}_T$ which is taken to be 0.4 in the current study \cite{stoll_porteagel_2006}. The diffusivities $\nu_T^{\text{SGS}}$ and $\kappa_T^{\text{SGS}}$ are modeled as \begin{align}
    \nu_T^{\text{SGS}}&=(C_s\tilde{\Delta})^2\sqrt{\tilde{S}_{ij}\tilde{S}_{ij}}\label{nuT_SGS},\mspace{50mu}
    \kappa_T^{\text{SGS}}=Pr_{\text{SGS}}^{-1}\nu_T^{\text{SGS}}=Pr_{\text{SGS}}^{-1}(C_s\tilde{\Delta})^2\sqrt{\tilde{S}_{ij}\tilde{S}_{ij}}. %\label{Pr_SGS},
\end{align}
where
$C_s$ is the Smagorinsky model coefficient and $\tilde{\Delta}=(\Delta x\Delta y \Delta z)^{1/3}$ is the filter width. The model coefficient $C_s$ is evaluated using the Lagrangian dynamic scale dependent model \cite{bouzeid2005}.  

The code uses the pseudo-spectral method along the streamwise and spanwise directions. The wall-normal direction is discretized using a second-order central finite difference method. The second-order accurate Adams-Bashforth scheme is used for time advancement. 
A shifted periodic boundary condition is used in the streamwise direction of the precursor domain to {\color{black}prevent artificially
long flow structures from developing. This approach also enables the development of statistical homogeneity with a shorter precursor domain size and less computational cost \cite{muntersetal2016}.}
% A shifted periodic boundary condition is used in the streamwise direction of the precursor domain to avoid the formation of {\color{black} elongated streamwise structures of excessive length that are known to develop as a result of streamwise periodic boundary conditions   \cite{muntersetal2016}.  In a finite box simulation with periodic boundaries,  these large  structures lead to spatial inhomogeneities in the computed statistics. Avoiding such structures without a shifted boundary condition would require a much longer domain which would drastically increase the computational cost. The advantage of using a shifted periodic condition is that it prevents artificially long flow  structures from developing and enabling to achieve statistical homogeneity with a shorter precursor domain size and less computational cost \cite{muntersetal2016}.}
% large coherent structures \cite{muntersetal2016}. {\color{black} In a finite box simulation with periodic boundaries, presence of these large coherent structures lead to statistical inhomogeneities. Resolving these structures will require a longer domain size which will drastically increase the computational cost. The advantage of using a shifted periodic condition is that it breaks these large turbulent structures enabling to achieve statistical homogeneity with a finite domain size and less computational cost \cite{muntersetal2016}.} 
A stress-free boundary condition is imposed on the top boundaries of the domains. The wall stress boundary condition from the equilibrium wall model
is applied at the bottom wall of both the precursor and wind turbine domains. Assuming the grid points near the surface is within the logarithmic layer (Monin-Obukhov similarity in the absence of stratification), the wall stress $\tau_w$ magnitude is evaluated as
\begin{align}
    \tau_w=-\left(\frac{\tilde{u}_r\kappa}{\ln(z_1/z_0)}\right)^2,\label{tau_wall_LES}
\end{align}
where $\tilde{u}_r=\sqrt{\tilde{u}^2+\tilde{v}^2}$ is the resultant horizontal velocity at the first grid point $z=z_1=\Delta z/2$, $\kappa=0.41$ is the Von Karman constant, and $z_0=0.1$ m is the assumed surface roughness height in the present study. Using equation \eqref{tau_wall_LES}, the wall stress components are evaluated as
\begin{align}
    \tau_{i,3|w}=\tau_w\frac{\tilde{u}_i}{\tilde{u}_r}, \ i=1,2,\label{tau_wall_i3_LES}
\end{align}
which are applied as the stress boundary conditions at the bottom boundary. For the CNBL simulation, we apply a zero buoyancy flux boundary condition $q_*=0$ to maintain the neutral stratification within the ABL.

The velocity fields of both the CNBL and TNBL LES are initialized with the log-law velocity profile  with zero-mean white noise superimposed. The noise is initialized for the entire domain in the TNBL while only for the first 900 m from the bottom surface for the CNBL.  To simulate the CNBL conditions in the LES, an initial potential temperature profile with a capping inversion layer is set up such that the boundary layer is neutral below the layer and is stably stratified above it (see figure \ref{schematic}). The capping inversion height is set to 1 km from the ground. The initial potential temperature ($\tilde{\theta}$) magnitude below the capping inversion region is 288 K (the same as the reference temperature scale $(\tilde{\theta}_0)$). The thickness of the capping inversion layer  where the potential temperature increases linearly from 288 K to 290.5 K is  100 m. Above this capping inversion layer, the potential temperature increases with a lapse rate of 0.001 K/m.

To represent the wind turbine, we use the local thrust coefficient-based actuator disk model (ADM) \citep{calaf_et_al_2010,shapiro_et_al_AD_2019,shapiro2020,meyers_meneveau_2010}. 
% {\color{black} We note that the Actuator Line Model (ALM) is a high-fidelity representation of the turbine as it can capture the effects of root and tip vortices behind the turbine which are not resolved by the ADM. However, owing to the high computational costs of running ALM and that there are not many differences in the far-wake behavior between ADM and ALM \cite{martinez_et_al_2012}, we choose ADM over ALM in this study.}  
The ADM treats the turbine as a drag disk of diameter $D$ and radius $R=D/2$ imparting a total force $(T)$ on the fluid {\color{black} directed along the unit normal direction $\bm n=\cos\beta\bm i+\sin\beta\bm j$ perpendicular to the disk (see figure \ref{schematic_veer})} given by
\begin{align}
    T=-\frac{1}{2}\rho_0\pi R^2C_T^\prime u_d^2 \label{Thrust},
\end{align}
where $C_T^\prime$ is the local thrust coefficient and  $u_d$ is the disk averaged velocity defined as
\begin{align}
    u_d=\int \tilde{\bm u}\cdot\bm n \ {\cal R}(\bm x) \ d^3\bm x \ {\color{black}=\int (\tilde{u}\cos\beta+\tilde{v}\sin\beta) \ \ {\cal R}(\bm x) \ d^3\bm x}. \label{ud}
\end{align}
{\color{black} The $u_d$ is an average of the velocity in the direction normal to the disk, $\tilde{\bm u}\cdot\bm n=\tilde{u}\cos\beta+\tilde{v}\sin\beta$}
% {\color{black} The $u_d$ is an average of the velocity along the normal direction $\tilde{\bm u}\cdot\bm n=\tilde{u}\cos\beta+\tilde{v}\sin\beta$} 
with the integration performed over the actuator disk using the indicator function ${\cal R}({\bf x})$ (defined below). The local thrust coefficient is set to $C_T^\prime= 1.33$, the same value as used in previous studies \citep{calaf_et_al_2010,shapiro_et_al_AD_2019,shapiro2020,meyers_meneveau_2010} to represent standard wind turbine operating conditions.
The force is spatially distributed using the smoothed indicator function $ {\cal R}(\bm x)$ such that the filtered force vector is
\begin{align}
    \tilde{\bm f}=T \, {\cal R}(\bm x)\,\bm n=T \ {\cal R}(\bm x)\cos\beta \ \bm i+ T \ R(\bm x)\sin\beta \ \bm j.\label{force_turbine}
\end{align}
The smoothed indicator function is defined according to \citep{LESGO,shapiro_et_al_AD_2019}
\begin{align}
    {\cal R}(\bm x)&=\int G(\bm x-\bm x^\prime) \ \mathcal{I}(\bm x^\prime) \ d^3\bm x^\prime,
\end{align}
where, $\mathcal{I}(\bm x)$ and $G(\bm x)$ are the normalized indicator function and Gaussian filtering kernel, respectively, given by
\begin{align}
    \mathcal{I}(\bm x)&=\frac{1}{s\pi R^2}\left[H(x+s/2)-H(x-s/2)\right]H(R-r),\label{ind_func_norm}\\
    G(\bm x)&=\left(\frac{6}{\pi \Delta ^2}\right)^{3/2}\exp\left(-6\frac{|\!|{\bm x}|\!|^2}{\Delta^2}\right).\label{Gauss_kernel}
\end{align}
In equation \eqref{ind_func_norm}, $s$ is the $x$-direction thickness of the forcing region 
%indicator function 
which is set to 10 m, $H(x)$ is the Heaviside function which is used to localize the disk within the region $-s/2<x<s/2$ and $r<R$, where $r=\sqrt{y^2+z^2}$. In the filtering kernel \eqref{Gauss_kernel}, $\Delta$ is the filter width defined as $\Delta=1.5 h$, where $h=\sqrt{\Delta x^2+\Delta y^2+\Delta z^2}$ is the effective grid size. 

{\color{black} We note that the Actuator Line Model (ALM) is a high-fidelity representation of the turbine as it can capture the effects of root and tip vortices behind the turbine which are not resolved by the ADM. However, owing to the high computational costs of running LES with ALM and that there are not many differences in the far-wake behavior between the ADM and ALM \cite{martinez_et_al_2012}, we choose ADM over ALM in this study.}

In the following sections, simulation setup and results from the precursor simulation of the CNBL and TNBL LES cases are discussed.

\subsection{Simulation setup\label{sim_setup_sec}}
\begin{figure}[H]
	\begin{center}
		\includegraphics[scale=.9]{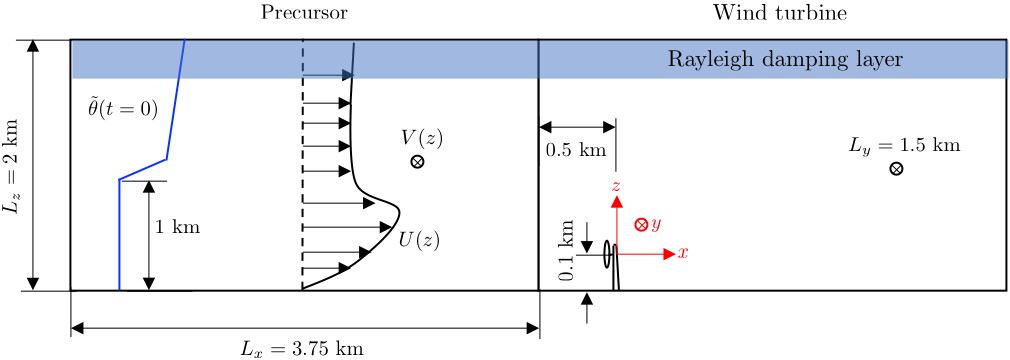}
	\end{center}
	\vspace*{-5mm}
	\caption{Schematic of the CNBL simulation setup with the turbine in the concurrent wind-turbine domain, inflow mean velocity profile with streamwise  $U(z)$ and wind veer $V(z)$ components, and initial potential temperature profile $\tilde{\theta}(t=0)$ in the precursor domain.} \label{schematic}
\end{figure}

%We use the concurrent-precursor method \cite{stevens2014} for performing the simulations. 
As shown in the schematic figure \ref{schematic}, the   simulation  is performed using two computational domains, the precursor and wind turbine domains. The turbine is placed in the wind turbine domain while the turbulent inflow simulating the CNBL or TNBL conditions is generated in the precursor domain. The details of the domain size and number of grid points are summarized in table \ref{tab:gridABL} and relevant dimensions are also shown in figure \ref{schematic}.

\begin{table}[H]
	\caption{\label{tab:gridABL} Computational domain size and grid points for LES of yawed wind turbine in the CNBL and TNBL cases. {\color{black} Note that the grid resolution is the same for both LES domains.}}
	\centering
	\begin{tabular}{cccc}
		\hline
		&&&\\
		Case&Precursor \& wind turbine&Number of grid points&Grid resolution\\
		&domain size&($N_x\times N_y\times N_z$)&($\Delta x (\text{m}) \times \Delta y  (\text{m}) \times \Delta z  (\text{m})$)\\
		&($L_x (\text{km}) \times L_y  (\text{km}) \times L_z  (\text{km})$)&&\\
		&&&\\ 
		CNBL&$3.75\times 1.5\times 2$ &$360\times 144\times 432$&\multirow{3}{*}{$10.4\times 10.4\times 4.6$}\\
		&&&\\
		TNBL&$3.75\times 1.5\times 1$&$360\times 144\times 216$&\\
		&&&\\ \hline
	\end{tabular}
\end{table}

Figure  \ref{schematic} includes a sketch of the initial potential temperature $(\tilde{\theta})$ profile (blue line) used to simulate the CNBL atmospheric condition in the precursor domain. A sponge (or Rayleigh damping) layer at the top boundary is used to dampen gravity waves in the computational domain. {\color{black} This is a wave absorbing layer of 500 m width from the top boundary. A body force with a cosine profile for its damping coefficient is applied in this layer to prevent reflection of gravity waves \citep{allaerts_meyers_2017,Durran_Klemp_1983}.} The resultant mean flow is expected to take the form of a low-level jet with velocity $\bm U=U(z)\bm i+V(z)\bm j$.
Together with the turbulence that develops in the precursor domain, the velocity  is then used as an inflow for the wind turbine domain {\color{black} using the concurrent-precursor method \cite{stevens2014}.  In this method, at each time step, a part of flow from the precursor domain is copied to the outflow region of the wind turbine domain. A fringe region is then defined to smoothly transition between the wind turbine flow and the region of flow copied from the precursor domain.}
The direction of the incoming mean velocity changes as a function of height. In contrast, for the TNBL simulation, the flow is isothermal in the entire computational domain and there is no wind veer, i.e., $V(z)\equiv 0$.

\begin{figure}[H]
	\begin{center}
		\includegraphics[scale=1]{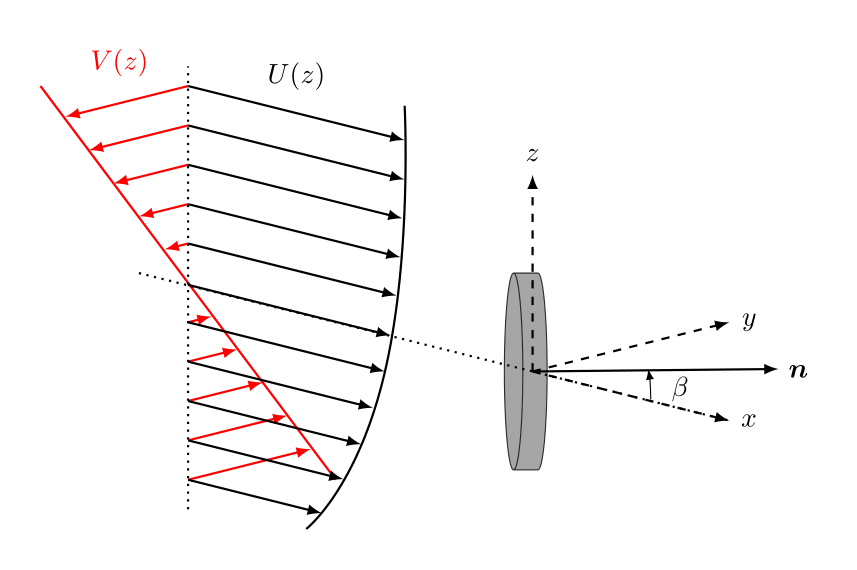}
	\end{center}
	\vspace*{-5mm}
	\caption{Schematic sketch of an isometric view of a yawed ($\beta$) wind turbine in the presence of ABL mean flow with velocity $\bm U=U(z)\ \bm i+V(z) \ \bm j$.} \label{schematic_veer}
\end{figure}

The wind turbine domain (second downstream domain in figure \ref{schematic}) includes a single turbine yawed at an angle $\beta=20^\circ$ placed at a distance of 500 m from the inlet. The diameter $(D)$ and hub height $(z_h)$ of the turbine are both 100 m. We use a Cartesian ($x\,y\,z$) coordinate system centered at the hub of the turbine where $x$, $y$, and $z$ are the streamwise, spanwise, and wall-normal directions, respectively. Figure \ref{schematic_veer} provides a sketch of the turbine orientation indicating the yaw angle $\beta$, the normal direction. Figure \ref{schematic_veer} also shows the two-component velocity profile that includes a veer velocity profile $V(z)$ in addition to the streamwise profile $U(z)$.

%This figure \dennice{do you mean figure 2, I do not see this in figure 1 and am not sure this sentence adds anything unless you want to describe figure 2 in more detail here. some suggested sentences follow.  Figure 2 provides a sketch of the turbine orientation indicating the veer angle $\beta$, the normal direction. This figure and the two component velocity profile that includes a veer velocity profile $V(z)$ in addition to the streamwise profile $U(z)$. }.  

\subsection{Characteristics of CNBL and TNBL \label{prec_result_sec}}

\begin{figure}[H]
	\begin{center}
		\includegraphics[scale=.8]{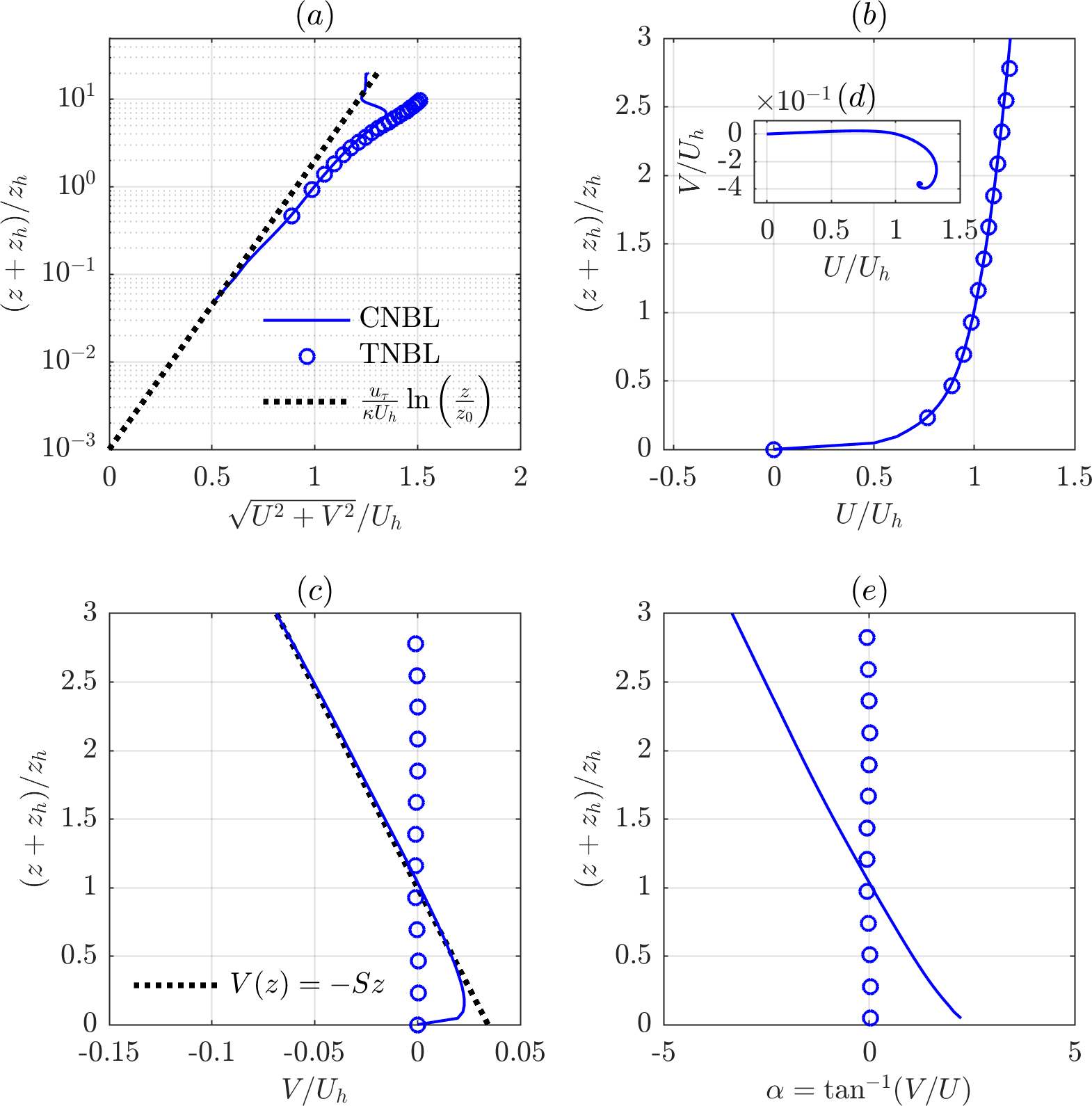}
	\end{center}
	\vspace*{-5mm}
	\caption{Plots of $(a)$ mean inflow velocity, $(b)$ streamwise velocity  and $(c)$ spanwise/wind veer velocity, $(d)$ Ekman spiral (inset in $(b)$), $(e)$ angle of inflow near hub height of the CNBL (\protect\blueline) and TNBL ({\color{blue} $\circ$})  LES cases.  Dotted black line in $(a)$ correspond to the Log-law velocity profile with $u_\tau=0.35$ m/s and $U_h=6.4$ m/s. The black dotted line in $(c)$ represents the linear approximation, $V(z)=-Sz$ with $S=2.2\times10^{-3} \ (1/s)$.} \label{mean_flow}
\end{figure}

Upon reaching statistically stationary conditions, mean velocity profiles are obtained from the LES in the precursor domain by performing time $(t)$ and horizontal spatial ($x$-$y$) averaging. 
Results are shown in figure \ref{mean_flow} 
for both the CNBL (blue solid line) and TNBL (open circles) cases. At the hub height, the magnitude of the mean velocity for the CNBL profile
%jet 
is $U_h=\sqrt{U^2(z=0)+V^2(z=0)}=6.4$ m/s, and the pressure gradient forcing for the TNBL is set so that the same $U_h=6.4$ m/s is implied for both simulations.
The friction velocities obtained from averaging the wall stress in the precursor domain of both cases are $u_\tau=0.35$ m/s. The resultant velocity profile for the CNBL and TNBL are shown in figure \ref{mean_flow}$(a)$ together with the log law velocity profile. Both LES cases have a similar mean wind profile within the turbulent boundary layer.
The streamwise $U(z)$ and  spanwise wind veer component $V(z)$ of the CNBL flow are plotted in figures \ref{mean_flow}$(b) \ \& \ (c)$. The normalized streamwise velocity $U/U_h$ from both the CNBL and TNBL cases agree quite well with each other. As expected, the TNBL case does not display any mean wind veer (blue circle markers in  \ref{mean_flow}$(c)$). Conversely, wind veer exists in the CNBL. This veer is well represented as $V(z)=-Sz$ in the region near the turbine hub height, i.e., covering the rotor region between $z=\pm 0.5D$. We find from LES the magnitude of the slope of the veer velocity is $S=2.2\times 10^{-3}$ (1/s). The Ekman spiral formed by $U(z)$ and $V(z)$ in the CNBL is shown in the inset of figure \ref{mean_flow}$(b)$. These results confirm that a streamwise aligned mean flow is achieved  at the hub height ($V(z=0)=0$) for the CNBL (figure \ref{schematic_veer}) using the proportional-integral (PI) controller approach \cite{sescu2014}.
This behavior is also evident in the plot of flow angle $\alpha(z)=\tan^{-1}[V(z)/U(z)]$ versus wall-normal height in figure \ref{mean_flow}$(e)$, where $\alpha(z=0)=0$. The corresponding plot (open circles) from the TNBL confirms that the flow is aligned in the streamwise direction throughout the  domain height. 
%\dennice{where is (f) I think you mean (e) and check that what I wrote is correct please}

\begin{figure}[H]
    \centering
    \includegraphics{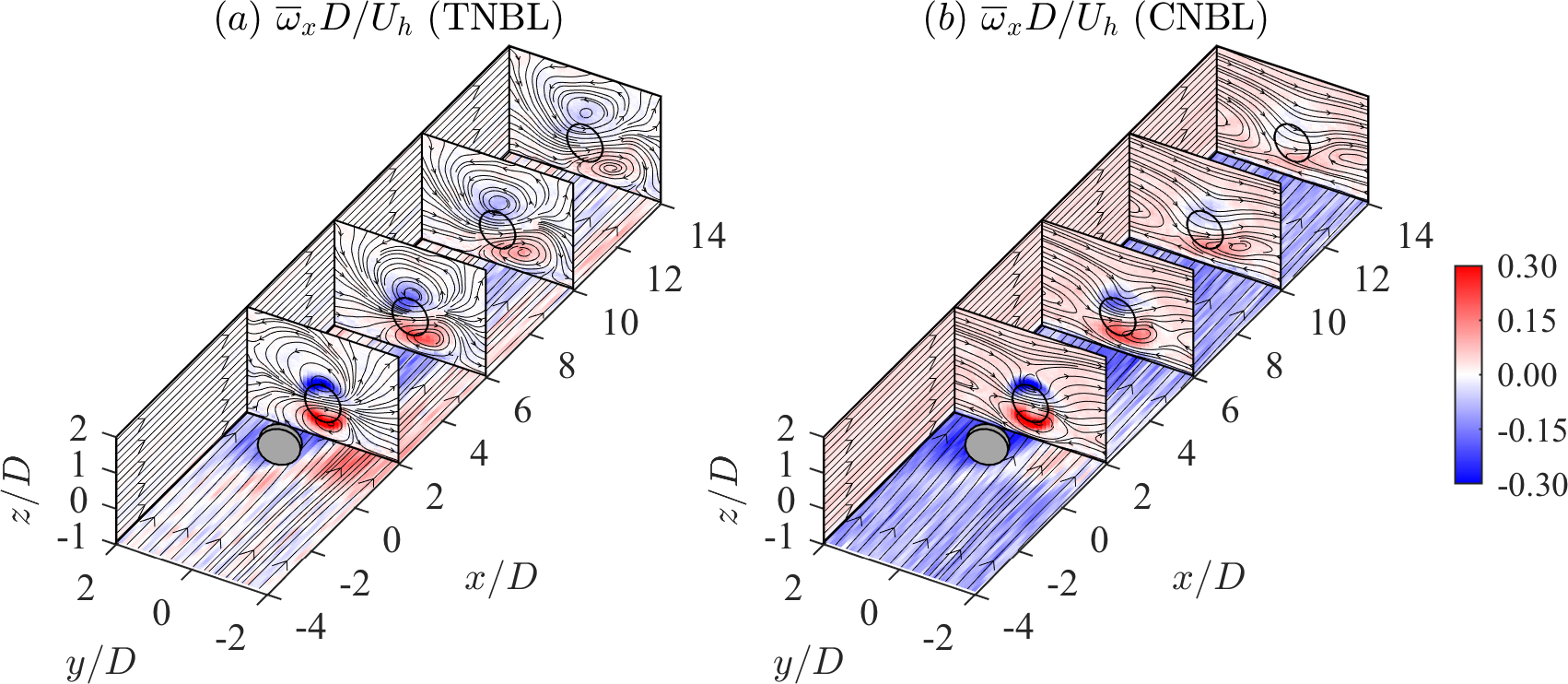}
    \caption{Contours  of time-averaged streamwise vorticity $\overline{\omega}_x$ behind a yawed ($\beta=20^\circ$) wind turbine from $(a)$ TNBL, and $(b)$ CNBL simulations. Contours are shown on various downstream planes at locations $x/D=[2,6,10,14]$ and on the bottom surface (first plane above the ground) and side plane at $y/D=2$.
    The vorticity contours are normalized by the turbine diameter $D$ and the hub height velocity $U_h$ of the incoming mean flow. The streamlines of the flow on the planes shown are overlayed on top of the contours visibly showing the presence of the cross-stream flow.}
    \label{cslice}
\end{figure}

% Figure \ref{cslice} shows contour plots of  instantaneous and time-averaged streamwise velocity for cross-stream planes at $x/D=[2,6,10,14]$ from the CNBL simulation.
% The streamlines overlayed on top of the contours on the cross-plane at $x/D=2$ indicate the presence of counter-rotating vortices in the wake region. There is a clearly visible clockwise rotating vortex below and a less visible counter-clockwise vortex on top. Similar results (not shown) are obtained for the TNBL case.
Figure \ref{cslice} shows contour plots of time-averaged streamwise vorticity $(\overline{\omega}_x)$ for cross-stream planes at $x/D=[2,6,10,14]$ from the TNBL and CNBL simulations.
The vorticity and the streamlines overlayed on top of the TNBL contours at the cross-planes indicate the presence of counter-rotating vortices in the wake region. There is a clearly visible clockwise rotating vortex below and counter-clockwise rotating vortex on top of the turbine as viewed along the positive $x$ direction in figure \ref{cslice}$(a)$. Whereas for the CNBL case in figure \ref{cslice}$(b)$, the contours show the presence of counter-rotating vortices affected significantly by the background veer vorticity where the top vortex is less visible than the bottom vortex.

\begin{figure}[H]
    \centering
    \includegraphics[scale=1]{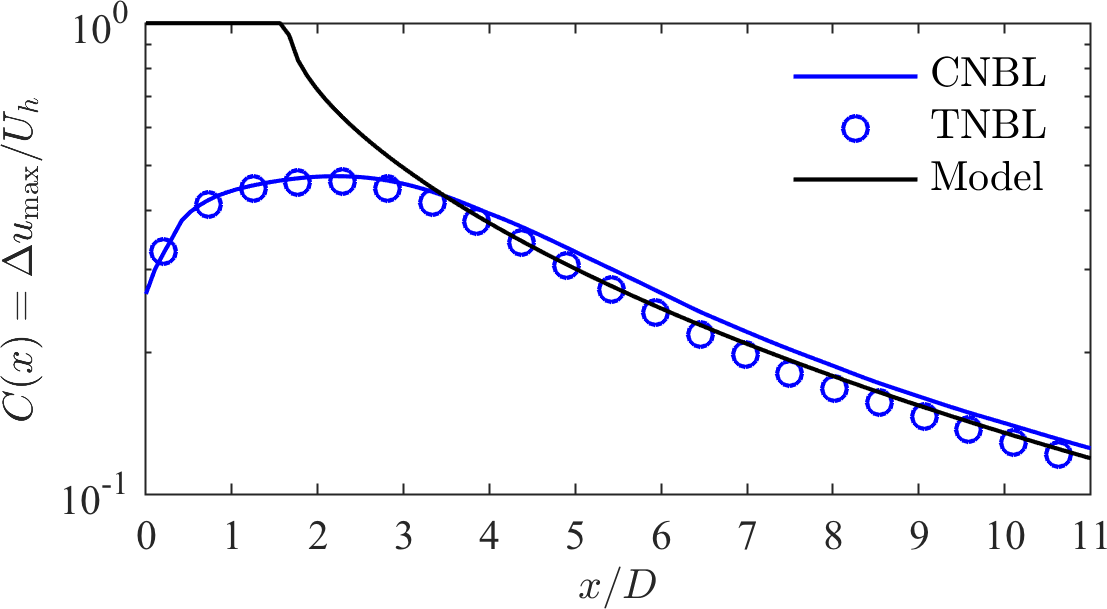}
  \caption{Streamwise evolution of $C(x)=\Delta u_{\mathrm{max}}/U_h$, the maximum velocity deficit normalized by $U_h$ in the wake of a yawed wind turbine for the CNBL case (solid blue line) and TNBL case (circles), compared to $C(x)$ from the model in equation \eqref{Cx} discussed in \S \ref{vortex_sheet_model} (black line; the model's validity holds farther downstream than the near rotor region (i.e., for $x/D>3$)).}
    \label{Cx_CNBL_TNBL}
\end{figure}

The decay of the wake strength is displayed in figure \ref{Cx_CNBL_TNBL}, which shows the maximum velocity deficit $(\Delta u_{\text{max}})$ normalized by $U_h$ in the wakes of the CNBL (blue line) and TNBL cases (circles) as a function of downstream distance from the turbine. Except for the near-rotor region ($x/D < 3$) they are very similar and also agree well with values obtained from a model (black line) to be discussed later in \S \ref{vortex_sheet_model}.
%\dennice{labeling is confusing here as the caption says $C(x)$ and the legend says sigma squared which is not defined anywhere}

% \begin{figure}[H]
%      \centering
%      \includegraphics{./Plots/du_contours_slice_CNBL}
%      \caption{Contours  of $(a)$ instantaneous, and $(b)$ time-averaged streamwise velocity behind a yawed ($\beta=20^\circ$) wind turbine from CNBL simulation. Contours are shown on various downstream planes at locations $x/D=[2,6,10,14]$ and on the bottom surface (first plane above the ground) and side plane at $y/D=2$.
%      The contours are normalized by the hub height velocity $U_h$ of the incoming mean flow. The streamlines of the flow on the planes shown are overlayed on top of the contours visibly showing the presence of the cross-stream flow.}
%      \label{cslice}
% \end{figure}
%{\bf ( Ghanesh: I feel here we need one nice visualization showing something with the turbine and the wake from LES, something that has the look of figure 8 of the Majid et al paper but should of course be different .. maybe a few yz planes with "sideways" v velocity contours showing how it gets more one sign above the turbines and another sign below? - not sure, but I think we need some such figure...)} 

In the next section, the time-averaged LES data is used to study the evolution and decay of the mean streamwise vorticity.
% Vortex Dynamics: 

\section{Generation and downstream mean vorticity evolution} \label{veer_vort_dyn}

%\dennice{Explain in the basic point/outline of this section in the intro paragraph.  The topic sentence I think does that but the focus of the paragraph seems to be the CVP model for the TNBL but that is not really what the section is about.}

In this section, we discuss the implications of including the wind veer effects on the model developed in Ref. \cite{shapiro2020}. The model is based on the linearized mean streamwise vorticity equation and includes two mechanisms: (1) Streamwise vorticity is generated at the turbine due to curl of the spanwise component of the thrust force, addressed in \S\ref{vort_Gen_sec}, and   (2) advection of the generated vorticity  downstream by the mean flow with simultaneous turbulent diffusion in the transverse $y$-$z$ plane, discussed in \S\ref{vort_decay_sec}.

\subsection{Mean streamwise vorticity generation at the yawed rotor\label{vort_Gen_sec}}

%\dennice{Introduce the model idea here before the equations}
%In this section, we review the theory of vorticity generation at the turbine location and investigate the implication of adding a wind veer vorticity. 

At the turbine location, the spanwise component of the   
force due to the yawed turbine changes along the spanwise direction. As a result, streamwise mean vorticity is created at the actuator disk periphery which is then transported downstream eventually forming the CVP.  At the rotor location, the linearized mean streamwise vorticity equation is dominated by advection and creation of vorticity due to the yawed actuator disk \citep{Tony_et_al_2017,shapiro2020}:   
%\dennice{is there a reference for this or a picture that would help clarify?}
\begin{align}
    U_h \frac{\partial  {\omega}_x}{\partial x}&=-\frac{1}{\rho_0}\frac{\partial  {f}_y}{\partial z},\label{vort_gen}
\end{align}
where ${\omega}_x$ represents the time-averaged streamwise vorticity
and ${f}_y$ is the spanwise ($y$-direction) force per unit volume exerted by the turbine rotor on the passing air. 

For a yawed actuator disk representation of the turbine,  the spanwise component of the thrust force per unit (fluid) mass  is given by \cite{shapiro2020}
\begin{align}
 {f}_y(x,y,z)=-{\frac{1}{2}\rho_0 C_T U^2_h\cos^2\beta \, H(R-r) \, \delta(x)} \ \sin\beta, \label{yaw_thrust} 
\end{align}
where $\beta$ is the yaw angle and $r^2=y^2+z^2$. Equation \eqref{yaw_thrust} is obtained using the standard expression for total thrust force   $T=-(1/2)\rho_0 \pi R^2 C_T U^2_h\cos^2\beta$ distributed uniformly on the actuator disk. This distribution is represented using the indicator function  ${\cal R}(\bm x)=H(R-r) \delta(x)/\pi R^2$ \cite{shapiro2018}, where
%In equation \eqref{yaw_thrust}, the indicator function $R(\bm x)=H(R-r)\delta(x)/\pi R^2$  is an approximation to the elliptic area of the projection of the yawed turbine diameter into the cross plane of the flow,  
$H(R-r)$ is the Heaviside function with the radial coordinate $r$ spanning the transverse $y$-$z$ plane with its origin at the hub of the turbine. The force acts only at the turbine's streamwise position $x=0$ as represented by the delta function $\delta(x)$. The thrust coefficient $C_T$ in equation \eqref{yaw_thrust} is related to the local thrust coefficient $C_T^\prime$ in equation \eqref{Thrust} by the relation $C_T=16C_T^\prime/(4+C_T^\prime\cos^2\beta)^2$ \cite{shapiro2018}. 

Using the expression \eqref{yaw_thrust} in \eqref{vort_gen}, the analytical form of the vorticity distribution generated at the turbine is obtained by 
integrating \eqref{vort_gen} in $x$ which gives
\begin{align}
 {\omega}_x(x=0,r,\theta)&=-\frac{1}{\rho_0 U_h}\int_{-\infty}^{x}\frac{\partial  {f}_y}{\partial z} \ dx=-\frac{1}{2} C_T U_h \ \cos^2\beta \ \sin\beta \ \sin\theta \  \delta(r-R) \ H(x).\label{omega_x_gen}
\end{align}
Here, the $r$ and $\theta$ directions are obtained through the cylindrical coordinate transformation of the $y$-$z$ transverse plane defined by $y=r\cos\theta$ and $z=r\sin\theta$. The corresponding circulation strength of the top and bottom vortices can be obtained by integrating equation \eqref{omega_x_gen} in the top and bottom half-planes respectively, which gives
\begin{align}
    \Gamma_{\text{top}}&=-\Gamma_{\text{bottom}}=\int_{0}^{\infty}\int_{0}^{\pi} {\omega}_x(x=0,r,\theta) \ r \ dr \ d\theta=-\Gamma_0, \label{circ_strength}
\end{align} 
where $\Gamma_0=R C_T U_h\cos^2\beta\sin\beta$ is the magnitude of the circulation strength.
The corresponding induced mean velocity components on the transverse plane ${v}$ and ${w}$ can be obtained by applying the Biot-Savart law \cite{shapiro2020} 
\begin{align}
    \begin{matrix}
     {v}&=\begin{cases}
-\dfrac{\Gamma_0}{4R}& r\leq R\\
&\\
-\dfrac{\Gamma_0}{4R}\dfrac{R^2}{r^2}\cos(2\theta)& r>R
\end{cases}
&
, ~~~~~~~~~~  {w}&=\begin{cases}
0&r\leq R\\
&\\
-\dfrac{\Gamma_0}{4R}\dfrac{R^2}{r^2}\sin(2\theta)& r>R
\end{cases}.
\end{matrix}\label{vw_estimate}
\end{align}

In order to compare the analytical solution  in equation \eqref{omega_x_gen} with LES, it must first be filtered at scales commensurate with the LES grid resolution to be consistent with the
filtered representation of the turbine forcing. This appropriate filtering was accomplished in Ref. \cite{shapiro2020} by first mapping the equation \eqref{omega_x_gen} to an arc distributed along the turbine edges (actuator disk circumference) and then convolving this mapped function  with a two-dimensional Gaussian filter.  
The resulting filtered vorticity at $x=0$ is given by
\begin{align}
    \omega_x(\theta,r)&=-\frac{\Gamma^*_0}{2R_*}\frac{\sin(\theta r/R_*)}{\sigma_R\sqrt{2\pi}} \exp\left(-\frac{(r-R_*)^2}{2\sigma_R^2}\right)\exp\left(-\frac{\sigma_R^2}{2R_*^2}\right), \label{omega_x_gen_numeq}
\end{align}
%\dennice{move to after the equation}
where, $R_*=R+0.75 h$ is the effective radius of the filtered actuator disk with $h$ depending on grid resolution according to $h=\sqrt{\Delta x^2+\Delta y^2+\Delta z^2}$. The Gaussian filter width is $\sigma_R=1.5 h /\sqrt{12}$. The magnitude of the effective circulation strength for the vortices expected from LES must be based on $R_*$ instead of $R$, specifically $\Gamma_0^*=R_*C_TU_h\cos^2\beta\sin\beta$.
Further following Ref. \cite{shapiro2020}, the analytical solutions for cross-stream induced velocities given by equation \eqref{vw_estimate} are compared with the numerical solutions by plotting $v+u_r\cos\theta$ and $w+u_r\sin\theta$. Here, $u_r$ is the radial velocity given by $u_r=(r/2)\partial_x u(R,0,0)$ for $r\leq R_*$ and $u_r=(R_*^2/2r)\partial_x u(R,0,0)$ for $r>R_*$. This step is done to eliminate the radial inflow that occurs due to streamtube expansion resulting from a finite streamwise pressure gradient behind the actuator disk that is not included in the theory but affects the LES results.  

\begin{figure}[H]
	\begin{center}
		\includegraphics[scale=.65]{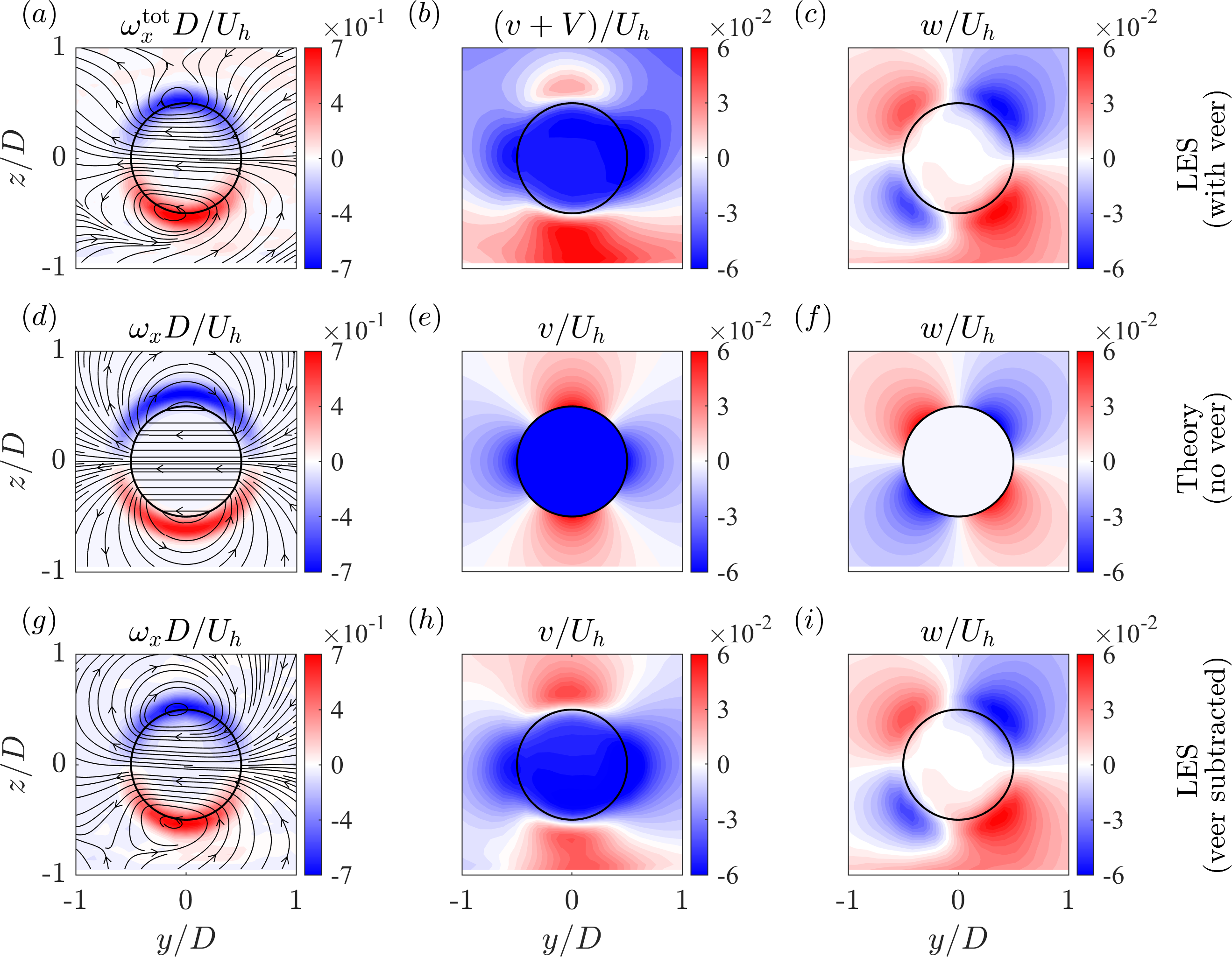}
	\end{center}
	\vspace*{-5mm}
	\caption{Contours of $\omega_x^{\mathrm{tot}} D/U_h$ (or) $\omega_x D/U_h$ $(a,d,g)$, $(v+V)/U_h$ (or) $v/U_h$ $(b,e,h)$ and $w/U_h$ $(c,f,i)$ at $x/D=0.5$ in the wake of yawed turbine with $C^\prime_T=1.33, \beta=20^\circ$.
	Plots ($a$-$c$) are from the CNBL LES, while ($d$-$f$) are analytical predictions   \cite{shapiro2020}.
	Subtracting the veer component of vorticity and spanwise velocity from  $(a,b,c)$ gives $(g,h,i)$, which has good agreement with the analytical estimates  $(d,e,f)$.  } \label{comparison_ctp1.33}
\end{figure}

With these considerations, we can now compare the induced velocities from LES including veer with the theory. The comparisons are shown in figures \ref{comparison_ctp1.33}.   
%, the LES results from the CNBL including the wind veer component are plotted while the veer is subtracted in figures \ref{comparison_ctp1.33}$(h)$-$(i)$. The wall-normal velocity [figures \ref{comparison_ctp1.33}$(c) \ \& \ (i)$] remain unaffected by the wind veer such that $\overline{w}=w$. The contours in figures \ref{comparison_ctp1.33}$(e)$-$(f)$ are from the analytical model for CVP in the absence of wind veer in a TNBL as discussed in Ref. \cite{shapiro2020}.  
Figure \ref{comparison_ctp1.33}(a) shows contours of the streamwise vorticity from the LES of the CNBL with streamlines of the cross-stream velocities superimposed. The corresponding filtered values from the analytical model (equation \eqref{omega_x_gen}) are shown in panel $(d)$.  
%\dennice{Based on the way we discuss this here I wonder about shifting the order of the panels}. 
%A comparison of these figures clearly illustrates both the structure of the CVP and the effect of wind veer on it.  
The LES results displayed in figure \ref{comparison_ctp1.33}$(a)$ show a significant asymmetry. Here, most of the streamlines at the top do not close around the top vortex, in contrast to those of the bottom vortex. This difference from the symmetric behavior predicted by the theory is due to the background mean vorticity caused by the veer, $ \Omega_x =-dV/dz$ and $ \Omega_y =dU/dz$.  The observation inspires a model in which the actual mean streamwise vorticity observed in LES (total vorticity ${\omega}^{\rm tot}_x$) comprises a linear superposition of the veer component of the vorticity ($\Omega_x$) and the yawed turbine-generated vorticity ($\omega_x$). 
\begin{align}
    {\omega}^{\rm tot}_x= {\omega_x}  +  \Omega_x. \label{vortx_veer}
\end{align}
We note that for a TNBL there is no wind veer $(V(z)=0)$ and the mean streamwise vorticity is $\Omega_x=0$.  
%which implies that the time-averaged vorticity $\overline{\omega}_x$ has contribution only from the CVP's vorticity $(\omega_x)$. 
If the assumed form in \eqref{vortx_veer} is correct, the difference ${\omega}^{\rm tot}_x - \Omega_x$ from LES (i.e.   subtracting the veer vorticity $\Omega_x =-dV/dz$) should be well predicted by the theoretical vorticity distribution from equation \eqref{omega_x_gen_numeq}. Figure \ref{comparison_ctp1.33}$(g)$  shows distributions of ${\omega}^{\rm tot}_x - \Omega_x$ from LES and its associated streamlines in the cross-plane. Clearly, the agreement between the veer-subtracted CNBL LES data and the model is much improved. For example, the symmetry is recovered and the streamlines now close around the top vortex as expected from the model. 

Figures \ref{comparison_ctp1.33}$(b),(e),(h)$ show the corresponding spanwise mean velocities, while panels $(c)$, $(f)$ and $(i)$ show the vertical velocities. Results in  figure \ref{comparison_ctp1.33}$(b)$   show significant asymmetry. Since  
%In summary, from the streamlines of the induced velocities (figures \ref{comparison_ctp1.33}$(a)\ \& \ (g)$), we can see that the veer relatively weakens the top vortex by distorting its structure significantly. 
%Since $V>0$ below the hub height, the time-averaged spanwise velocity $ {v}^{\rm tot}=v+V(z)$ (figure \ref{comparison_ctp1.33}$(b)$) is predominantly positive for $z/D<0$. 
$V>0$ below the hub height and $V<0$ above the hub height, the induced velocity above the rotor disk is significantly lower in the LES than in the model. In LES the veer velocity cancels much of the induced velocity from the yawed turbine.   Upon subtracting the veer velocity distribution  from  the CNBL LES results (figures \ref{comparison_ctp1.33}$(g)$-$(i)$), we retrieve the more symmetric distribution, with good agreement with the theory (figures \ref{comparison_ctp1.33}$(d)$-$(f)$). Contours obtained from the LES of the TNBL case (not shown) also show good agreement with the model.
%and TNBL contours (figures \ref{comparison_ctp1.33}$(j)$-$(l)$). 

These results suggest that the model proposed in Ref. \cite{shapiro2020} to predict the generation of mean streamwise vorticity due to a yawed wind turbine remains essentially unaltered even in the presence of a wind veer in a CNBL. The background veer vorticity can be added to that of the unperturbed vorticity distribution in order to model the total vorticity distribution immediately behind the yawed actuator disk. The further downstream evolution of mean vorticity is discussed in the following sub-section.

\subsection{Downstream evolution of mean vorticity}
\label{vort_decay_sec}
Downstream of the rotor, the streamwise evolution of the turbine-generated streamwise mean vorticity ${\omega}_x$ is dominated by mean advection and transverse turbulent diffusion \cite{shapiro2020},
\begin{align}
    U_h\frac{\partial {\omega}_x}{\partial x}&=\nu_T(x)\left(\frac{\partial^2 {\omega}_x}{\partial y^2}+\frac{\partial^2 {\omega}_x}{\partial z^2}\right).\label{vort_decay}
\end{align}
Here, $\nu_T(x)\sim u_\tau l(x)$ is the turbulent eddy viscosity, which can be modeled as the product of the friction velocity $u_\tau$ and a mixing length scale $l(x)$.
%\dennice{It is more clear to just write the equation}. 
The wake size, which increases linearly downstream of the turbine in the presence of ambient turbulence, is chosen as the appropriate mixing length scale, such that $l(x)\sim x$ \cite{shapiroetal2019}. Following Ref. \citep{shapiro2020}, this linearly growing mixing length scale is given by
$l(x)=2\kappa_\nu(x-x_0)/\sqrt{24}$, where $k_\nu=u_\tau/U_h$ is the vortex expansion coefficient and $x_0=-24^{-1/4}1.5 h/k_\nu$ is the virtual origin. Using this $l(x)$, the eddy viscosity $\nu_T(x)$ becomes
\begin{align}
    \nu_T(x)=u_\tau 2\kappa_\nu(x-x_0)/\sqrt{24}.\label{nuT_model}
\end{align}
From the current LES simulations, using the values $u_\tau=0.35$ m/s and $U_h=6.4$ m/s, the vortex expansion coefficient for both simulations results in $k_\nu=0.0547$.

We note that adding the constant vorticity due to veer, 
$\Omega_x$ %\dennice{that is not clear here as there are derivitives in all three directions and in fact you discuss other assumptiosn needed next please clarify} 
to ${\omega}_x$, does not affect equation \eqref{vort_decay} %\dennice{put the number for clarity} 
as long as it is constant in the vertical direction. Also, from figure \ref{mean_flow}, we can see that near the turbine, the magnitude of the veer velocity ($V(z)$) is much smaller than the streamwise mean flow $(U(z))$. Therefore, we can expect the streamwise advection term to be dominant resulting in the same governing equation \eqref{vort_gen} for $\omega_x$ even in the presence of veer in a CNBL near the hub height. We note that advection of vorticity by the veer-affected mean velocity can be simply accounted for by interpreting $x$ above as the streamline coordinate along the vortex trajectory that has a speed $V(z)$ in the transverse direction, but for now, we shall neglect these differences.

The solution to the advection-diffusion equation \eqref{vort_decay} can be obtained by the linear superposition of the fundamental solution to the equation. The vorticity generated at the turbine can be regarded as a distribution of multiple point vortices with vorticity magnitude $d{\omega}_x(x=0,y,z)=d\Gamma_p \ \delta(y-y_0) \ \delta(z-z_0)$ of circulation strength $d\Gamma_p$ located at the turbine disk edges $y_0=R\cos\theta,z_0=R\sin\theta$.
Governed by equation \eqref{vort_decay}, these point vortices are  advected downstream by the mean flow with simultaneous turbulent diffusion along the transverse directions. Correspondingly, the fundamental solution \cite{saffman_1993} to equation \eqref{vort_decay} is 
\begin{align}
    d {\omega}_x(x,y,z)=-\frac{d\Gamma_p}{4\pi^2\eta^2(x)}\exp\left(-\frac{(y-R\cos\theta)^2+(z-R\sin\theta)^2}{4\eta^2(x)}\right),\label{funda_soln}
\end{align}
where $\eta(x)$ is the transverse turbulent diffusion length scale \cite{shapiroetal2019} given by
\begin{align}
    \eta(x)=\sqrt{\frac{1}{U_h}\int_{0}^{x}\nu_T(\xi) \  d\xi}. \label{visc}
\end{align}
Using the eddy viscosity from \eqref{nuT_model} in \eqref{visc}, $\eta(x)$ becomes
\begin{align}
    \eta(x)=k_\nu(x-x_0)/24^{1/4}. \label{eta_knu}
 \end{align}

Adding the contributions from all the point vortices distributed on the turbine edge, the resultant downstream vorticity field is obtained. Adding this vorticity to that of the background veer leads to a total vorticity of
\begin{align}
    {\omega}^{\rm tot}_x(x,y,z) = -\int_{0}^{2\pi} \frac{d\Gamma_p}{4\pi^2\eta^2(x)} \ \exp\left(-\frac{(y-R\cos\theta)^2+(z-R\sin\theta)^2}{4\eta^2(x)}\right) + \Omega_x.\label{omegax_downstream_CVP}
\end{align}
We assume the peak vorticity occurs at $y=0$ and $z=\pm R$, evaluating \eqref{omegax_downstream_CVP} at this location gives \cite{shapiro2020}
\begin{align}
     {\omega}^{\rm tot}_{x,max}&=\frac{\Gamma_0}{4\eta^2(x)}\exp\left(-\frac{R^2}{2\eta^2(x)}\right)I_1\left(\frac{R^2}{2\eta^2(x)}\right) + \Omega_x,\label{omegax_max}
\end{align}
where $I_n$ is the $n^{\text{th}}$ order modified Bessel function of the first kind.

The circulation strength $\Gamma(x)$ of the generated vorticity (excluding the background veer vorticity) as a function of the downstream location is obtained by integrating $\omega_x$ in the transverse plane, $\Gamma(x)=|\int_{0}^{\infty}\int_{-\infty}^{\infty}{\omega}_x(x,y,z) \ dy \ dz|=|\int_{-\infty}^{0}\int_{-\infty}^{\infty}{\omega}_x(x,y,z) \ dy \ dz|$, which yields \cite{shapiro2020}
\begin{align}
    \frac{\Gamma(x)}{\Gamma_0}&=\frac{\sqrt{\pi}}{4}\frac{R}{\eta(x)}\exp\left(-\frac{R^2}{8\eta^2(x)}\right)\left[I_0\left(\frac{R^2}{8\eta^2(x)}\right)+I_1\left(\frac{R^2}{8\eta^2(x)}\right)\right].\label{circ_decay}
\end{align}
These analytical predictions for the decay of the maximum vorticity and circulation strength were shown to have good agreement with the numerical simulations in Ref. \cite{shapiro2020}.

\begin{figure}[H]
	\begin{center}
		\includegraphics[scale=.96]{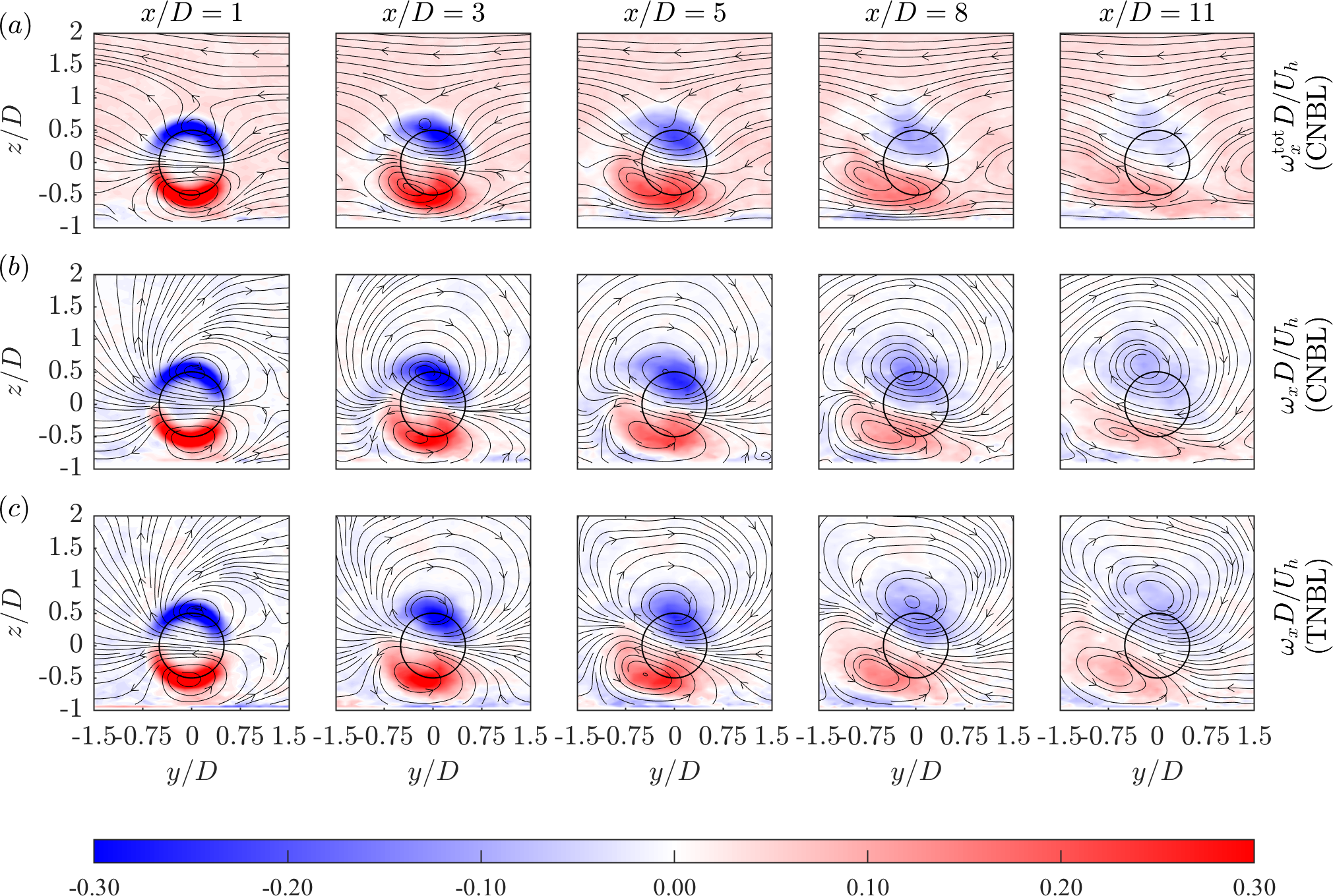}
	\end{center}
	\vspace*{-5mm}
	\caption{Turbulent decay of streamwise vorticity from the CNBL $(a)$ with background veer, $(b)$ after subtracting veer  component and from $(c)$ TNBL case at downstream locations $x/D=1,3,5,8,11$.} \label{wx_veer_no_veer}
\end{figure}

In figure \ref{wx_veer_no_veer}, the time-averaged streamwise vorticity distribution from the CNBL and TNBL simulations are plotted at different downstream locations, here cross-stream velocity streamlines are superimposed. The results from the CNBL LES (i.e., including wind veer in figure \ref{wx_veer_no_veer}$(a)$) can be compared to those with the veer subtracted  (figure \ref{wx_veer_no_veer}$(b)$) and the contours from the TNBL simulation (figure \ref{wx_veer_no_veer}$(c)$). Similar to prior results at the turbine location,  it is evident from the cross-stream velocity streamlines that veer distorts the structure of the top and bottom vortices  along their entire downstream evolution. This distortion results in an asymmetric distribution of vorticity between the top and bottom vortices. 

Next, we compare the top and bottom vortex circulations from LES to the model. 
The circulation in LES is computed by integrating total vorticity 
$\Gamma_{\rm LES}(x)=|\int_A{\omega}^{\rm tot}_x(x,y,z) \ dy \ dz|$ over a suitably defined cross-sectional area $A$ of the vortices. Following the approach of Ref. \cite{shapiro2020}, we use Otsu's classification method to determine the cross-sectional region over which to integrate the vorticity. We apply the calculation of circulation using both the full vorticity 
%$\overline{\omega}_x$ from LES, 
and after subtracting the background veer vorticity. 
%, i.e. integrating only
%$\overline{\omega}_x-\langle {\omega}_x \rangle$.

\begin{figure}[H]
	\begin{center}
		\includegraphics[scale=1]{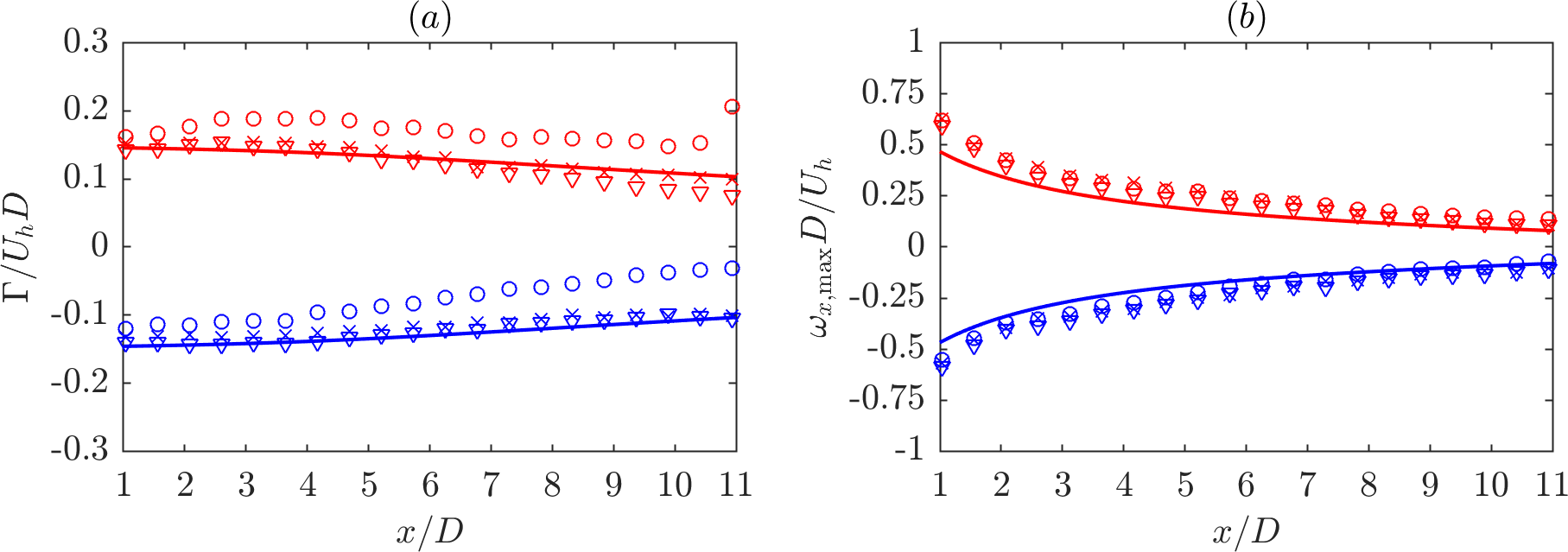}
	\end{center}
	\vspace*{-5mm}
	\caption{Decay of $(a)$ circulation strength $\Gamma/U_h D$ as a function of the downstream location and $(b)$ maximum streamwise vorticity  $\omega_{x,\text{max}}D/U_h$.  Red and blue colors correspond to  bottom and top vortices of the CVP respectively. Solid lines (\protect\redline,\protect\blueline) represent analytical solutions \eqref{omegax_max} and \eqref{circ_decay}. Markers represent quantities plotted from LES, where circular markers (\tikzcircle[red]{2pt},\tikzcircle[blue]{2pt}) include wind veer component and open triangular markers ({\color{red}$\triangledown$},{\color{blue}$\triangledown$}) are plotted after subtracting the wind veer component, ({\color{red} $\times$},{\color{blue} $\times$}) are plotted from the TNBL simulation.} \label{wxmax_circ}
\end{figure}

The resulting decay of circulation for the CNBL case is shown in  figure \ref{wxmax_circ}$(a)$, where the unequal strength of the top (blue circles) and bottom (red circles) vortices can be seen. Veer causes the circulation strength of the bottom vortex to be significantly higher than the top vortex's circulation. Also, these circulation estimates deviate from the analytical prediction of equation \eqref{circ_decay} (solid lines in the plot). However, upon subtracting the veer vorticity $\Omega_x$ from the time-averaged vorticity (open triangles), 
we obtain good agreement  with the theoretical estimate from $\eqref{circ_decay}$ 
{\color{black} with small differences at far downstream locations for the bottom vortex}. It also shows good agreement with the circulation strengths obtained from the TNBL simulation (cross markers). Figure \ref{wxmax_circ}$(b)$ shows that the peak vorticity of the CVP ($\omega_{x,\text{max}}$) is  much larger than the veer vorticity $(\Omega_x)$ so that whether the $\Omega_x$ is included or not, does not affect the results appreciably.

{\color{black} In figure \ref{wxmax_circ}$(a)$, we note a small deviation of the red open triangles at far downstream locations. This difference can be attributed to the size difference of the bottom vortex in the CNBL versus the TNBL cases (see figures \ref{wx_veer_no_veer}$(b)$ \& $(c)$ at $x/D=8,11$). The difference could be due to a differing turbulent expansion rate of the vortices  between the CNBL and TNBL cases. The presence of a capping inversion layer and the stable Geostrophic region above a CNBL are known to limit the size of the largest turbulent eddies \citep{Kitaigorodskii_Joffre_1988,wyngaard_2010}. %This may lead to the availability of relatively lesser turbulence within CNBL than TNBL. 
Thus, turbulent diffusion may be more effective in the case of a TNBL, thus enlarging the bottom vortex more when compared to the CNBL.
%leading to dissimilar vortex sizes. 
While the bottom vortices grow slightly differently in the TNBL and CNBL cases, the peak vorticity strength at downstream locations from figure \ref{wxmax_circ}$(b)$ are quite comparable. Hence, the circulation strength of the bottom vortex in a CNBL falls slightly below the analytical model due to its smaller size. }

The comparisons of LES results to the theoretical model confirm that for the presently studied CNBL case the downstream evolution of mean vorticity remains dynamically unaffected by the presence of wind veer. The only effect of the veer is its mean  vorticity is superimposed with that of the vorticity generated by the yawed actuator disk. With improved understanding and an analytical model for the downstream evolution of vorticity in the wake of a yawed turbine, in the next section, we examine analytical modeling of the mean axial velocity deficit needed to predict the power generation of downstream wind turbines.

\section{Application of vortex sheet-based curled wake model in a CNBL}
\label{vortex_sheet_model}

The discussions in the previous sections demonstrate that a yawed actuator disk in the presence of an incoming mean flow sheds a pair of vortices behind the turbine. This CVP  arises from an initially  tubular vortex sheet distribution generated around the turbine disk periphery \citep{Martinez_Tossas_2020}.
%\dennice{Does this need a reference that does not come from the previous discussions}.  
The induced velocity from the vortex sheet deflects the wake away from the centerline and also deforms the vortex sheet, leading to the known curled wake structure \cite{howland16}.
%\dennice{cite Howland here?}. 
A recent study \cite{bastankhah_et_al_2022}, proposes an analytical model for the downstream evolution of vorticity that reproduces the curled wake shape from the induced velocities acting on the vortex sheet. We first summarize the model of \cite{bastankhah_et_al_2022} in \S\ref{sec:vortex_mdl_summary}. We then  generalize it to include the new insights regarding possible effects of veer from the previous section and compare it with results from LES in \S\ref{veer_CNBL_sec}. 

\subsection{Summary of vortex sheet-based curled wake model}
\label{sec:vortex_mdl_summary}
According to the vortex sheet-based curled wake model, the streamwise velocity deficit  $\Delta u(x,y,z)=U(z)-u(x,y,z)$ is given by a Gaussian profile \citep{bastankhah2014,bastankhah2016} according to 
\begin{align*}
    \frac{\Delta u}{U_h}&=C(x)\exp\left[-\frac{(y-y_c)^2+(z-z_h)^2}{2\,\sigma(\theta,x)^2}\right].\numberthis\label{du_majid}
\end{align*}
Here, $C(x)$ is the magnitude of the normalized velocity deficit, while $\sigma(\theta,x)$ is the wake width expressed as a function of the downstream location $x$ and polar angle $\theta$ defined as $\tan\theta=(z-z_h)/(y-y_c)$. The centroid location of the wake is $y_c,z_h$. In this section, we shift the origin of the $x\,y\,z$ coordinate system  from the center of the turbine to the ground in order to be consistent with the curled wake model definitions in Ref. \cite{bastankhah_et_al_2022}. 

\begin{figure}[H]
	\begin{center}
		\includegraphics[scale=1]{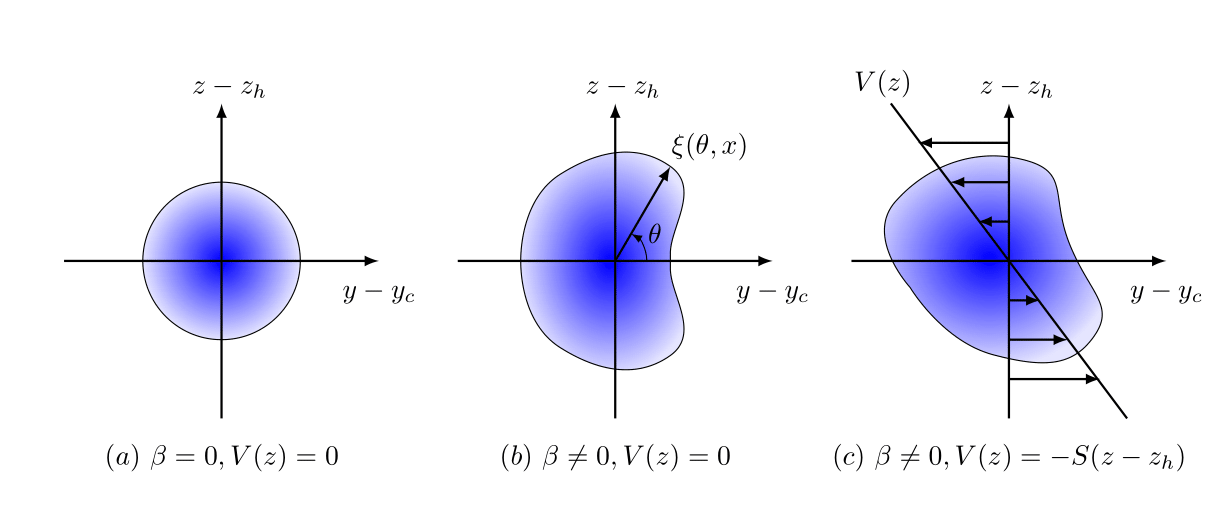}
	\end{center}
	\vspace*{-5mm}
	\caption{Sketch of velocity deficit behind $(a)$ an unyawed turbine with no wind veer, $(b)$ a yawed turbine showing the induced curled wake, with the angle-dependent wake shape function $\xi(\theta,x)=\xi_0(\theta)\, \hat{\xi}(\theta,\hat{t})$ and no wind veer, and $(c)$ the yawed turbine in a CNBL (with wind veer). The shading represents the strength of the velocity deficit $\Delta u$.} \label{sketch1}
\end{figure}

%\subsubsection{Wake radius $\sigma(\theta,x)$}
The wake width $\sigma(x,\theta)$ is determined by adding Jensen model linear wake growth \cite{Jensen_1983} and the angle-dependent wake shape function according to 
\begin{align*}
    \sigma(\theta,x)&=k_w \, x+0.4 \, \xi(\theta,x) \numberthis\label{sigma_width}.
\end{align*}
Here $k_w=0.6 u_\tau/U_h$ is the wake expansion coefficient \cite{bastankhah_et_al_2022}, and the curled structure of the wake is modeled by the angle-dependent wake shape function  $\xi(\theta,x)$. %This function describes the vortex sheet location including the variation of the wake width along the polar angle $\theta$ at any given streamwise location. 
This function describes the vortex sheet location which augments the wake width along the polar angle $\theta$ at any given streamwise location as per equation \eqref{sigma_width}.
%\dennice{this could use clarification} 
The sketches in Figure \ref{sketch1} show the shape of the wake behind the turbine under different conditions. Figure \ref{sketch1}$(a)$ shows the symmetric Gaussian wake of an unyawed turbine that expands linearly downstream of the flow in a TNBL without wind veer effects. Figure \ref{sketch1}$(b)$  shows the definition of the angle-dependent %radius of the wake 
wake shape function
$\xi(\theta,x)$ which becomes curled due to the vortex sheet self-induction again in the absence of wind veer. Figure \ref{sketch1}$(c)$ shows the sketch of the wake behind a yawed turbine in an ABL with wind veer effects, which is further discussed in  \S\ref{veer_CNBL_sec}.
The model in \cite{bastankhah_et_al_2022} also takes into account the effect of the hub vortex which occurs as a result of the rotation of the turbine. However, in this study, we {\color{black} do not} include the effects of rotation.

The wake shape function $\xi(\theta,x)$ has units of length and is normalized using $\xi_0(\theta)=\xi(\theta,0)$, the initial shape of the wake which depends on the angle $\theta$ and represents the shape of an ellipse as the rotor disk is yawed
%(depending on angle $\theta$ as an ellipse if the rotor disk is yawed)
%\dennice{you lost me here}) 
\begin{align}
    \xi_0(\theta)=R\sqrt{A_*}\frac{|\cos\beta|}{\sqrt{1-\sin^2\beta\sin^2\theta}} \label{xi0_theta}.
\end{align}
Here $A_*$ is the ratio of the expanded streamtube area with the projected frontal area of the rotor \cite{bastankhah_et_al_2022}:
\begin{align}
    A_*=\frac{1+\sqrt{1-C_T\cos^2\beta}}{2\sqrt{1-C_T \cos^2\beta}}.\label{A_star} 
\end{align}
The normalization introduces a dimensionless angle-dependent wake shape function 
$\hat{\xi}(\theta,\hat{t})$ according to 
\begin{align}
\xi(\theta,x) =  \xi_0(\theta) \, \hat{\xi}(\theta,\hat{t}).
\end{align}
where $\hat{t}$ is a dimensionless time variable to be defined below \cite{bastankhah_et_al_2022}.

The downstream evolution  of the angle-dependent radial location of the wake is given by $\xi(\theta,x)=\xi_0(\theta)+\int u_r(\theta,t) \ dt$, where time $t$ stands for the downstream evolution, i.e., $t=x/U_h$, and $u_r(\theta,t)$ is the radial velocity induced by the shed streamwise vorticity.   The vortex sheet shape at downstream locations %\dennice{for a given $\theta$?} 
for a given $\theta$
is changed by the induced radial velocity from the vortices.  Given the strength of the vortex sheet or the circulation density $\gamma=\gamma(\theta,t)$, the radial velocity $u_r(\theta,t)$ governing the shape of the vortex sheet can be determined using the Biot-Savart law as described in Ref. \cite{bastankhah_et_al_2022}. 
In Ref. \cite{bastankhah_et_al_2022}, the evolution of dimensionless velocities, $\hat{u}_r=u_r/\gamma_b$, $\hat{u}_\theta=u_\theta/\gamma_b$ and sheet location, $\hat{\xi}=\xi/\xi_0$ are obtained using a power series method, where   $\gamma_b$ is the reference circulation density at the turbine location. It is related to the circulation strength $\Gamma_0$ in equation \eqref{circ_strength} by
\begin{align}
    \gamma_b=\frac{\Gamma_0}{2R}=\frac{1}{2}C_T U_h \cos^2\beta \sin \beta.
\end{align}
  The dimensionless time is defined as $\hat{t}=t \, \gamma_b/\xi_0$.   The analytical general solutions are valid at short times, i.e. 
only near the turbine 
for times ($|\hat{t}|\leq 2$) \cite{bastankhah_et_al_2022}.

To capture the wake-curling effects at large times, an empirical expression for the non-dimensional vortex sheet location $\hat{\xi}(\theta,\hat{t})$ is proposed (see equation B1 in Ref. \cite{bastankhah_et_al_2022}) which is valid for both short and long times. 
For a non-rotating turbine, the empirical expression simplifies to
\begin{align*}
\hat{\xi}(\theta,\hat{t})&=1-\alpha\biggl[
\frac{1}{2}\tanh\left(\frac{\hat{t}^{2}}{4\alpha}\right)\cos2\theta
-\frac{1}{4}\tanh\left(\frac{\hat{t}^{3}}{8\alpha}\right)\cos(3\theta)\\
&\mspace{50mu}
-\frac{5}{48}\tanh\left(\frac{\hat{t}^{4}}{16\alpha}\right)\cos(2\theta)
+\frac{7}{48}\tanh\left(\frac{\hat{t}^{4}}{16\alpha}\right)\cos(4\theta)  \biggr],\numberthis\label{xi_emp}
\end{align*}
where, the constant  $\alpha=1.263$. 
In equation \eqref{xi_emp}, we also need to know $\hat{t}$. 
%and \cite{bastankhah_et_al_2022} propose a model for the dimensionless time too. 
In a turbulent boundary layer, the strength of the vortex decays downstream leading to a time-dependent circulation density $\gamma_b=\gamma_b(t)$ \cite{bastankhah_et_al_2022}. The definition of the non-dimensional time $\hat{t}=t\gamma_b/\xi_0$ then generalizes to
\begin{align}
    \hat{t}=\frac{1}{\xi_0}\int_{0}^{t}\gamma_b(t^\prime) \ dt^\prime. \label{that_CVP_int}
\end{align}
The circulation density $\gamma_b(t)$ is related to the decaying circulation strength $\Gamma(x)$ considered in equation
\eqref{circ_decay} by $\gamma_b=\Gamma(x)/2R$. Using this relation along with $t \approx x/U(z)$ with $U(z)=(u_\tau/\kappa) \ln(z/z_0)$ and $\eta(x)=k_\nu x/24^{1/4}$ from equation \eqref{eta_knu}, and assuming the virtual origin $x_0=0$ to simplify the problem, the integral \eqref{that_CVP_int} becomes

\begin{align}
    \hat{t}=\frac{24^{1/4}}{2k_\nu U(z)\xi_0 R}\int_{0}^{\eta}\Gamma_b(\eta^\prime) \ d\eta^\prime.\label{that_Gamma}
\end{align}
%The integral was evaluated numerically in Ref. \cite{bastankhah_et_al_2022} and using a fitting function for the integral, the approximation for $\hat{t}$ is expressed compactly according to  
%\dennice{why is the numerical part mentioned I am a bit confused}
The integral was approximated using a fitting function resulting in a  $\hat{t}$ which is expressed compactly according to
\begin{align}
    \hat{t}(x,z)&\approx-1.44\frac{U_h}{u_\tau}\frac{R}{R\sqrt{A_*}}C_T\cos^2\beta\sin\beta\left[1-\exp\left(-0.35\frac{u_\tau}{U(z)}\frac{x}{R}\right)\right].\label{that}
\end{align}
The value of $\hat{t}$ from equation \eqref{that} is used to evaluate $\hat{\xi}(\theta,\hat{t})$ in \eqref{xi_emp} and subsequently obtain $\xi(\theta,x)=\xi_0(\theta) \hat{\xi}(\theta,\hat{t})$ in equation \eqref{sigma_width}.

%\subsubsection{Model for normalized velocity deficit magnitude $C(x)$}
Next, we turn to the normalized maximum velocity deficit magnitude $C(x)=\Delta u_{\text{max}}/U_h$
in equation \eqref{du_majid} which is given by
\begin{align}
C(x)&=1-\sqrt{1-\frac{C_T\cos^3\beta}{2\tilde{\sigma}^2(x)/R^2}},\label{Cx}
\end{align}
where, 
\begin{align}
    \tilde{\sigma}^2(x)=(k_w  x+0.4 R\sqrt{A_*})(k_w  x+0.4 R\sqrt{A_*}\cos\beta), \label{sigma_tilde}
\end{align} 
%is an estimate of the average wake width over the angles,
%\dennice{?? you mean capturs this trend or something?} and allows the horizontal width to be reduced due to yawing at an angle $\beta$.
is an estimate of the average wake width over the angles which captures the trend of reduction of horizontal width due to yawing at an angle $\beta$.
The quantity $R\sqrt{A_*}$ is also an approximation of the initial wake shape according to equation \eqref{xi0_theta} for small $\beta$.

In figure \ref{Cx_CNBL_TNBL}, this analytical prediction for the maximum normalized velocity deficit magnitude $C(x)$ is  plotted alongside estimates from the CNBL and TNBL LES. Here it is clear that the model provides good agreement with the LES and is able to predict the correct  velocity deficit decay for $x/D\gtrsim 3$.

%\subsubsection{Wake deflection estimate from curled wake model}

The curled wake model predicts the wake deflection $y_c$ due to the shed vorticity according to 
\begin{align}
    y_c&=\hat{y}_c(\hat{t}) \ R\sqrt{A_*},\label{yc_majid}
\end{align}
where, the non-dimensional wake deflection $\hat{y}_c$ is expressed as
\begin{align}
    \hat{y}_c(\hat{t})&=\frac{(\pi-1)|\hat{t}|^3+2\sqrt{3}\pi^2\hat{t}^2+48(\pi-1)^2|\hat{t}|}{2\pi(\pi-1)\hat{t}^2+4\sqrt{3}\pi^2|\hat{t}|+96(\pi-1)^2}\text{sgn}(\hat{t})-\frac{2}{\pi}\frac{\hat{t}}{[(z+z_h)/R\sqrt{A_*}]^2-1}.\label{yhatc}
\end{align}
In Ref. \cite{bastankhah_et_al_2022}, an analytical solution for the wake deflection for short times is obtained assuming a circular vortex sheet. For longer times, the deflection of the wake is estimated instead by modeling the vorticity distribution as a CVP. This first term in equation \eqref{yhatc} which is an empirical fit captures the behavior of both these solutions across short and long times.
The second term models the deflection due to the image vortices needed to model ground effects.  

\begin{figure}[H]
    \centering
    \includegraphics[scale=1]{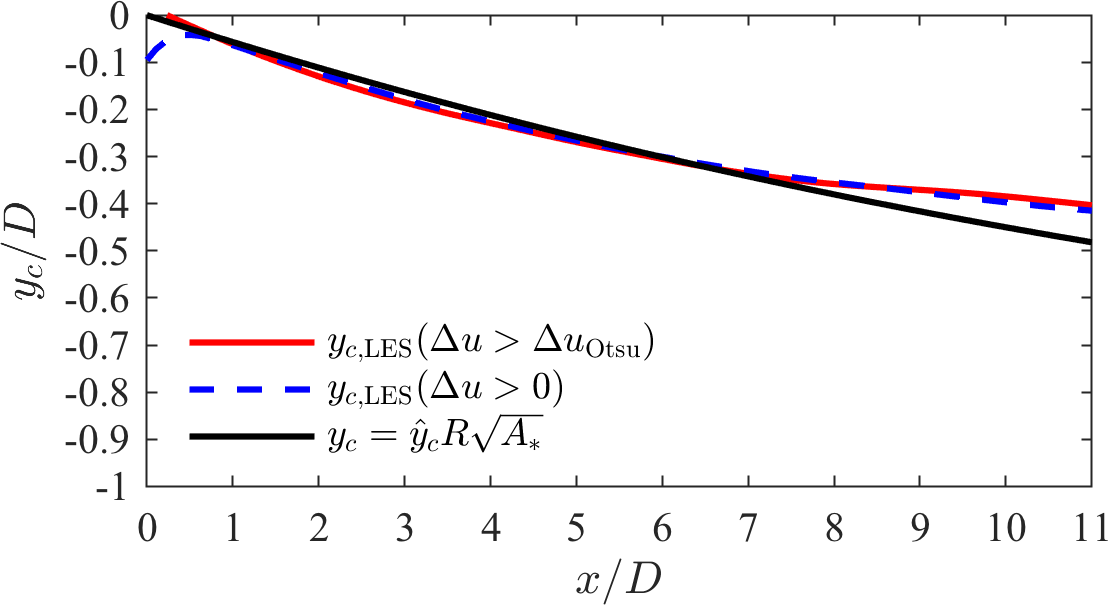}
    \caption{Wake deflection at hub height from the LES (CNBL) determined using two different methods is compared with the estimate from curled wake model (\protect\blackline) \cite{bastankhah_et_al_2022}. The LES results include $y_{c,\mathrm{LES}}$ using the integration over positive velocity defect region (\protect\bluedashedline) or Otsu's method (\protect\redline).
    %and $y_{c,\mathrm{LES},2D}$ also using the integration over positive velocity defect region (\protect\bluedotline) or Otsu's method (\protect\reddashdotline).
    }
    \label{yc_CNBL_TNBL}
\end{figure}

These analytical estimates for wake deflection in equation \eqref{yc_majid} are compared with the wake deflection computed from the velocity deficit in the CNBL simulation in figure \ref{yc_CNBL_TNBL}. The two lines in color  represent the wake deflection  $y_{c,\text{LES}}$ obtained from the CNBL simulation using
%the following two methods,
\begin{align}
    y_{c,\text{LES}}(x)&=\frac{\int_y y \ \Delta u(x,y,z_h) \ dy}{\int_y \Delta u(x,y,z_h) \ dy}, % ~~~~~~~
    %y_{c,\text{LES},2D}(x) =\frac{\int_y\int_z y \ \Delta u(x,y,z) \ dy \ dz}{\int_y\int_z \Delta u(x,y,z) \ dy \ dz},
\label{yc_LES_12D}
\end{align}
where $\Delta u(x,y,z)=U(z)-u(x,y,z)$ is the velocity deficit. We  chose two methods, one in which regions where $\Delta u>0$ and another where $\Delta u>\Delta u_{\text{Otsu}}$ is used to perform the integrations in equation \eqref{yc_LES_12D}. Here, $\Delta u_{\text{Otsu}}$ is the threshold chosen by the Otsu edge detection method \cite{shapiro2020}. 
%Wake deflection from CNBL simulation evaluated using these two methods are shown in figure \ref{yc_CNBL_TNBL}. 
It is evident in figure \ref{yc_CNBL_TNBL} that the wake deflection from the curled wake model given by equation \eqref{yc_majid} shows good agreement with the LES up to $x/D=7$, but slightly overpredicts the observed deflection further downstream. For the purposes of the present modeling accuracy, however, we consider the level of agreement satisfactory.

\subsection{Curled wake model applied to the  CNBL including veer}
\label{veer_CNBL_sec}
%Since wind veer exists in a CNBL, we also need to add its effect on the shape of the velocity deficit. 
Following Ref. \cite{Abkar_et_al_2018}, the effect of veer can be included as an additional spanwise wake displacement such that
\begin{align}
    y_{c,\text{Veer}}(x,z) =\frac{x}{U(z)} \, V(z) \approx  -\frac{x}{U(z)}\,S \, (z-z_h). \label{yc_veer}
\end{align}
Here we also take into account the additional effects from $z$-dependent streamwise velocity. This effect induces larger sideways displacement in the lower parts of the domain since for a fixed $x$, the advection time will be larger and the sideways displacement will have had more ``time'' to take place.  The total spanwise wake deflection of the velocity deficit in \eqref{du_majid} including the veer displacement \eqref{yc_veer} is 
\begin{align}
y_c(x,z) = \hat{y}_c(\hat{t}) \, R \sqrt{A_*} \, - \, \frac{x}{U(z)} \,S \, (z-z_h).
  \label{yc_CVP_veer}
\end{align}
where, $\hat{y}_c(\hat{t})$ is the dimensionless deflection given by  equation \eqref{yhatc} with $\hat{t}=\hat{t}(x,z)$  evaluated from equation \eqref{that}.  For the present LES data, the shear rate $S$ of the wind veer in equation \eqref{yc_veer} has been measured, and results in $S=2.2\times10^{-3}$ (1/s)  (see figure \ref{mean_flow}$(c)$).

\begin{figure}[H]
    \centering
    \includegraphics[scale=1]{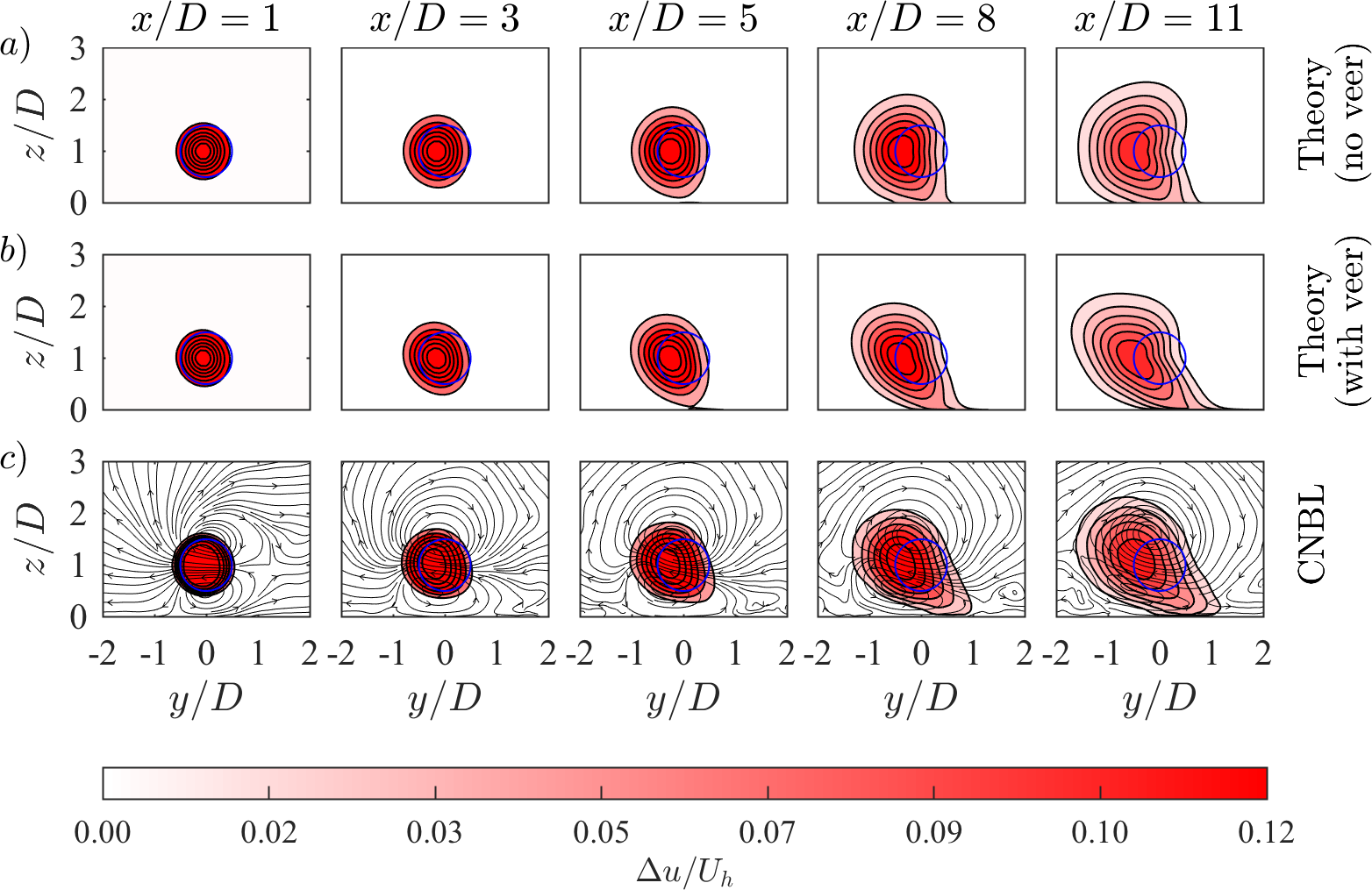}
    \caption{Contours of $\Delta u/U_h$ at various streamwise locations from curled wake analytical model \eqref{du_majid}  with centroid location $(a)$ $y_c=y_{c,1}$, $(b)$ $y_c=y_{c,1}+y_{c,\text{Veer}}$, and $(c)$ CNBL simulation with the $v$-$w$ velocity streamlines.}
    \label{du_CNBL_analytic}
\end{figure}

\begin{figure}[H]
    \centering
    \includegraphics[scale=.9]{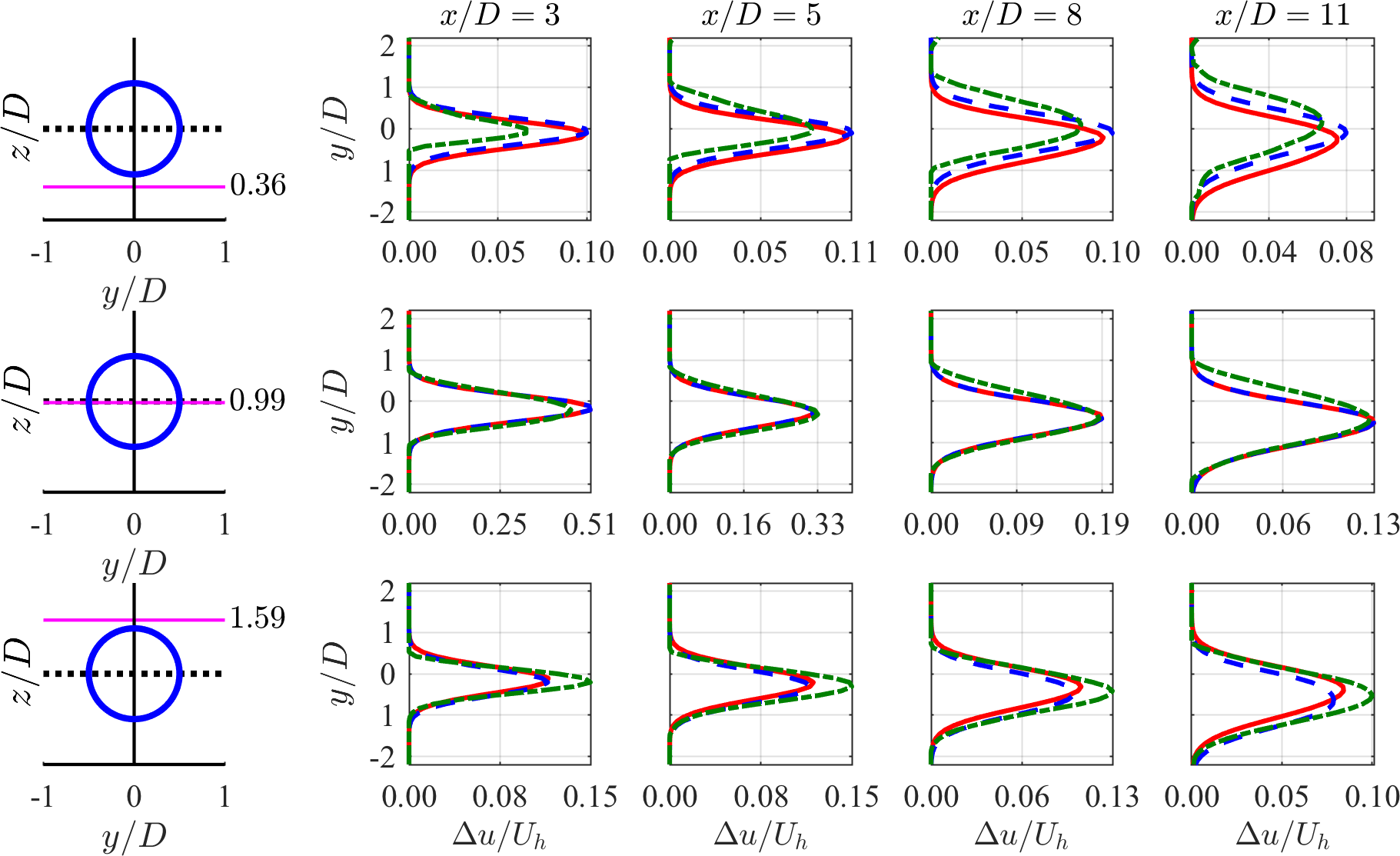}
    \caption{Profiles of $\Delta u/U_h$ plotted at $z/D=[0.36,1,1.59]$ and different streamwise locations $x/D=[3,5,8,11]$ from curled wake analytical model \eqref{du_majid} with centroid location  $y_c=y_{c,1}$ (\protect \redline), $y_c=y_{c,1}+y_{c,\text{Veer}}$ (\protect \bluedashedline), and CNBL simulation (\protect \greendashdotline).}
    \label{du_CNBL_analytic_y1D}
\end{figure}

\begin{figure}[H]
    \centering
    \includegraphics[scale=.9]{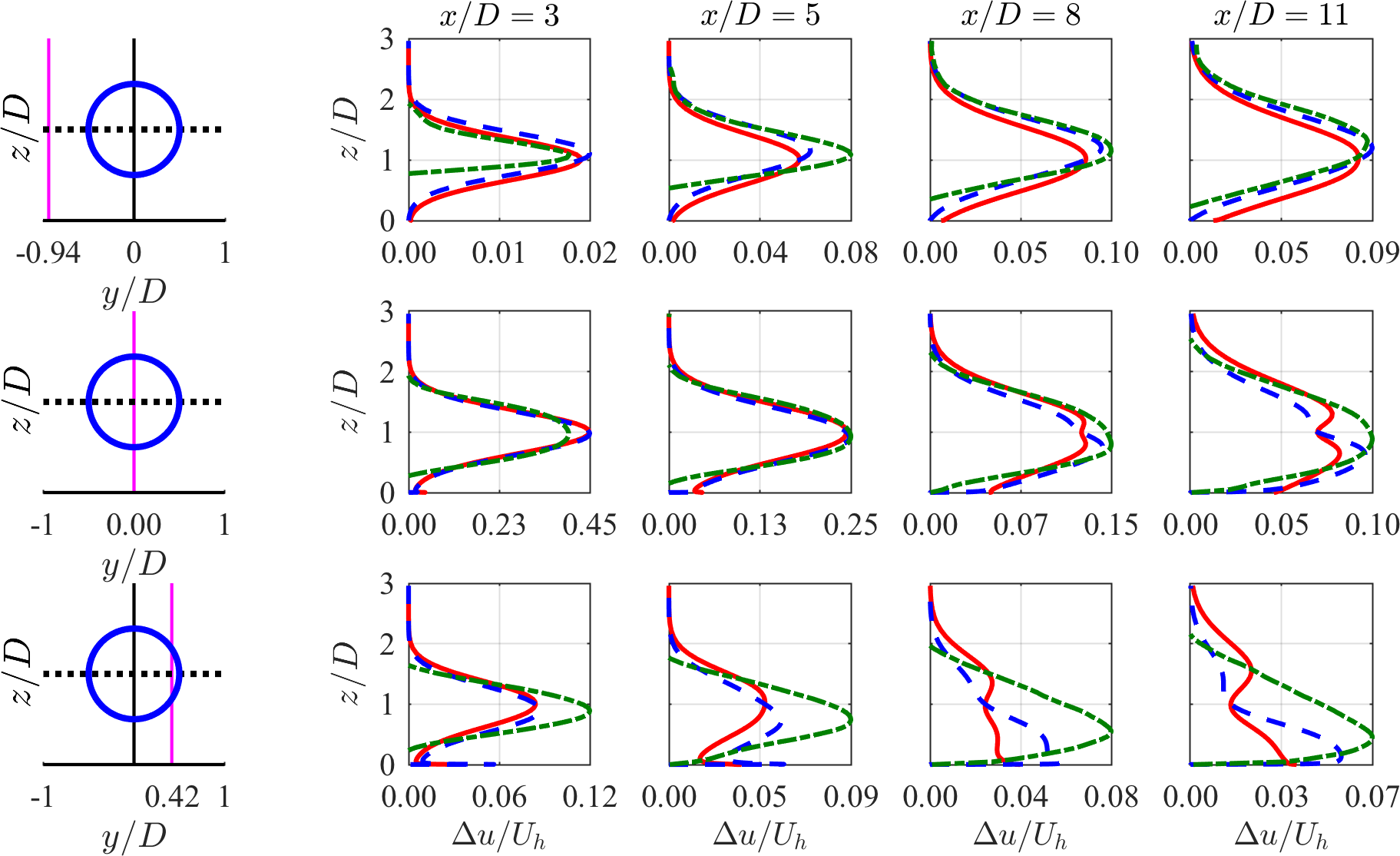}
    \caption{Profiles of $\Delta u/U_h$ plotted at $y/D=[-0.94,0,0.42]$ and different streamwise locations $x/D=[3,5,8,11]$ from curled wake analytical model \eqref{du_majid} with centroid location  $y_c=y_{c,1}$ (\protect \redline), $y_c=y_{c,1}+y_{c,\text{Veer}}$ (\protect \bluedashedline), and CNBL simulation (\protect \greendashdotline).}
    \label{du_CNBL_analytic_z1D}
\end{figure}

Analytical model predictions for $\Delta u/U_h$ from equation \eqref{du_majid} at select downstream distances  are plotted as contours in figure \ref{du_CNBL_analytic}$(a)$ (no veer), \ref{du_CNBL_analytic}$(b)$ (including veer) and compared against the CNBL data (figure \ref{du_CNBL_analytic}$(c)$). Panel (c) also includes  $v$-$w$ cross-stream velocity streamlines (evaluated from LES with the veer velocity subtracted). {\color{black}
In figures 
\ref{du_CNBL_analytic_y1D} and \ref{du_CNBL_analytic_z1D}, 
%12 and 13
the one-dimensional profiles of $\Delta u/U_h$ at $z/D=[0.36,1,1.59]$ and $y/D=[-0.94,0,0.42]$ respectively are plotted at different downstream locations.} We can clearly see the effect of the wind veer in the contours of the analytically modeled defect velocity. The veer deflects the wake  rightward below the hub height and leftward above the hub height, whereas not including the veer produces a wake that is only deformed by the vortex-sheet streamwise vorticity, resulting in a curled wake structure. The analytically predicted contours  closely resemble the contours from the CNBL simulation. {\color{black} The one-dimensional profiles in figures 
\ref{du_CNBL_analytic_y1D} and \ref{du_CNBL_analytic_z1D}
%12 and 13 
show a more detailed comparison between the model and LES. These plots show that the veer-deflected velocity deficit structure at far downstream locations is better predicted by the wake model with the veer correction term included, albeit with slight deviations.}

% Analytical model predictions for $\Delta u/U_h$ from equation \eqref{du_majid} at select downstream distances  are plotted as contours in figure \ref{du_CNBL_analytic}$(a)$ (no veer), \ref{du_CNBL_analytic}$(b)$ (including veer) and compared against CNBL (figure \ref{du_CNBL_analytic}$(c)$). Panel (c) also includes  $v$-$w$ cross-stream velocity streamlines (evaluated from LES with the veer velocity subtracted). {\color{black}
% In figures \ref{du_CNBL_analytic_y1D} and \ref{du_CNBL_analytic_z1D}, the one-dimensional profiles of $\Delta u/U_h$ at $z/D=[0.36,1,1.59]$ and $y/D=[-0.94,0,0.42]$ respectively are plotted at different downstream locations.} We can clearly see the effect of the wind veer in the contours of the analytically modeled defect velocity. The veer deflects the wake  rightward below the hub height and leftward above the hub height, whereas not including the veer produces a wake that is only deformed by the vortex-sheet streamwise vorticity, resulting in a curled wake structure. The analytically predicted contours  closely resemble the contours from the CNBL simulation. {\color{black} The one-dimensional profiles from figures \ref{du_CNBL_analytic_y1D} \& \ref{du_CNBL_analytic_z1D} show a more detailed the precise comparison between the model and LES. The plots show that the veer-deflected velocity deficit structure at far downstream locations is better predicted by the wake model with the veer correction term included, albeit with slight deviations.}

The results suggest that the curled wake model which was originally developed for a TNBL can also be applied to a CNBL. The presence of veer only alters the shape of the velocity deficit by shifting it towards one side below the hub height and to the other side above the hub height, and this effect can be modeled with the additional spanwise deflection term given by equation \eqref{yc_veer}.

\begin{figure}[H]
    \centering
    \includegraphics[scale=1]{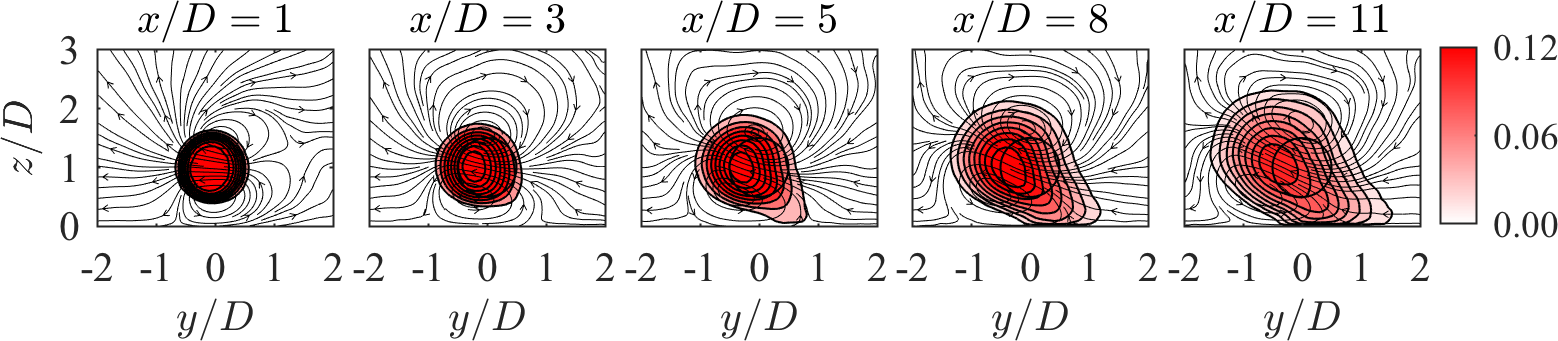}
    \caption{Contours of $\Delta u/U_h$ from the TNBL simulation with cross-stream $v$-$w$ velocity streamlines.}
    \label{du_TNBL}
\end{figure}

To complete the description of velocity defect evolution,  we compare with results from the TNBL simulation in figure \ref{du_TNBL}, again  including the cross-stream velocity streamlines. Although there is no wind veer in this simulation, we observe that the wake structure is still deflected to the right below the hub height and somewhat towards the left above the hub height. This may appear  surprising at first sight since one would expect that the CVP induces a symmetric curled wake structure akin to the analytical contours from figure \ref{du_CNBL_analytic}$(a)$ without veer. However, in this case, the cause for the asymmetry  appears to be that unlike what is assumed in the model, the top and bottom vortices themselves are not aligned in the same spanwise location. We can clearly see from the contour at $x/D=11$ in  figure \ref{du_TNBL} that the bottom vortex has moved relatively more towards the left than the top vortex. The cause of the misalignment of the two vortices is likely due to the fact that the streamwise advection velocity near the ground is smaller than that at the top due to the shear in the $U(z)$ ABL profile. Thus, at a fixed distance $x$ from the turbine, moving with the mean velocity there has been more time for a sideways motion for the bottom vortex (induced by the top vortex) as compared to the top vortex (induced by the bottom vortex). As a result, the entire bottom vortex is shifted to the left, hence deflecting the induced velocity near the wake center downwards. This effect is partially accounted for by using $U(z)$ in equation \eqref{that} in the model and hence also helps explain the slight top-bottom asymmetry seen in figure \ref{du_CNBL_analytic}(a) even in the absence of veer. However, the results show that for the TNBL case this effect is somewhat stronger. More detailed model versions that take such effects into account may have to be developed. However, the presently proposed approach already provides significant accuracy in predicting the highly non-trivial spatial distribution of velocity in yawed turbine wakes. 

\section{Conclusions}\label{conclusion}
This study investigates the effect of wind veer on the evolution of mean streamwise vorticity and velocity deficit in the wake of a yawed turbine, in the context of conventionally neutral atmospheric boundary layers including Coriolis acceleration causing a wind veer. In earlier work in the absence of veer (TNBL), the curled wake dynamics had been shown to be best understood as consisting  of a vortex generation phase, where streamwise vorticity is injected in the flow due to the curl of the turbine yaw force, followed by downstream advection and transverse turbulent diffusion of vorticity.  Analytical solutions determining such behavior were derived and validated in Ref. \cite{shapiro2020}. In the present study, we have shown that the evolution of mean vorticity remains relatively unchanged even in the presence of wind veer. LES data show that when wind veer is included, the streamwise vorticity corresponding to the veering mean flow, $\Omega_x$, is simply superimposed additively to the vorticity generated by the yawed turbine. For the yaw angle considered in this study, this strengthens the bottom vortex and weakens the top vortex  while their evolution is still modeled well following the approach as described in Ref. \cite{shapiro2020}. We also adapted the vortex sheet-based curled wake model \cite{bastankhah_et_al_2022} to describe present flow conditions. It was found to yield good results for modeling the velocity deficit in both conventionally-neutral and truly-neutral conditions. The velocity deficit affects downstream wind turbines if interacting with the wake.
%The model is able to predict the right decay of the velocity deficit magnitude and wake deflection at the hub height. 
The presence of wind veer in the CNBL can be included in the wake model by adding an additional spanwise displacement term, modeled here using the wind veer velocity $V(z)$. We approximated the latter as $V(z)=-S (z-z_h)$, where  the shear rate parameter $S$ was determined from the LES of CNBL flow. 
{\color{black} Although only one yaw angle is discussed in this work, the veer correction can in principle be applied to model the evolution of the velocity deficit for different yaw angles, at least up to those tested in the work of \cite{bastankhah_et_al_2022}, i.e., up to $\beta = 30^0$.}
% {\color{black} Although only one yaw angle is discussed in this work, the veer correction can in principle be applied to model velocity deficit for different yaw angles, at least up to those tested in the work of \cite{bastankhah_et_al_2022}, i.e., up to $\beta = 30^0$. The additional veer correction can still be applied to the wake model to simulate the effect of the veer. }
{\color{black} Also, note that both the LES and the model proposed do not include the effects of wake angular momentum. Such effects can be included in the analytical model as well as in the LES using ADM with rotation (ADM-R), as was done in \cite{bastankhah_et_al_2022}.  A detailed analysis of such additional effects, i.e., evaluating whether wake angular momentum could affect the superposition of veer and yaw vorticity decay, is left for future efforts.
}
% {\color{black} Also, note that both the LES and the model proposed do not include the effects of wake angular momentum. For cases where wake rotation is significant, we remark that the LES of 
% \cite{bastankhah_et_al_2022} included ADM with rotation (ADM-R) and the model included wake deformation due to rotation. Thus it is possible to include such effects in the present model, but we refrain from speculating whether wake angular momentum could further affect the superposition of veer and yaw vorticity decay.
% }
For a fully predictive analytical model, one requires, in addition, an analytical model for $V(z)$ and $U(z)$. Further work is required to cast such models (e.g. those developed in Refs. \citep{krishna_ellison_1980,roisin_malacic_1997}) in terms that can be efficiently incorporated into present wind turbine wake models.
\section*{Acknowledgments}
This research has been supported by the National Science Foundation (grant CBET-1949778). We are thankful to the National Center for Atmospheric Research (NCAR) for providing the computational resources. Discussions with Dr. Carl Shapiro are greatly appreciated. 
\bibliography{biblio}% Produces the bibliography vica BibTeX.
\end{document}